\documentclass[aps,prx,fixfloat,twocolumn]{revtex4-2}
\usepackage{graphicx,amsmath,amsfonts,bm, color}
\usepackage{wasysym}

\usepackage{xcolor,hyperref}
\hypersetup{
   colorlinks,
   linkcolor={blue!50!black},
   citecolor={blue!50!black},
   urlcolor={blue!80!black}
}

\usepackage{enumitem}

\renewcommand{\=}{\!=\!}
\newcommand{\1}{^{\mbox{\tiny (1)}}}

\newcommand{\dbar}{{\,\mathchar'26\mkern-12mu d}}
\DeclareMathAlphabet{\mathitbf}{OML}{cmm}{b}{it}

\definecolor{darkGreen}{RGB}{4,161,85}

\begin{document}

\title{Experimental evidence for the $\omega^4$ tail of the nonphononic spectra of glasses}
\author{Avraham Moriel$^{1}$}
\author{Edan Lerner$^{2}$}
\author{Eran Bouchbinder$^{3}$}
\email{eran.bouchbinder@weizmann.ac.il}
\affiliation{$^{1}$Dept.~of Mechanical \& Aerospace Engineering, Princeton University, Princeton, New Jersey 08544, USA\\
$^{2}$Institute for Theoretical Physics, University of Amsterdam, Science Park 904, Amsterdam, Netherlands\\
$^{3}$Chemical \& Biological Physics Dept., Weizmann Institute of Science, Rehovot 7610001, Israel}

\begin{abstract}
It is now established that glasses feature low-frequency, nonphononic excitations, in addition to phonons that follow Debye's vibrational density of state (VDoS). Extensive computer studies demonstrated that these nonphononic, glassy excitations follow a universal non-Debye VDoS ${\cal D}_{\rm G}(\omega)\!\sim\!\omega^4$, at low frequencies $\omega$. Yet, due to intrinsic difficulties in disentangling ${\cal D}_{\rm G}(\omega)$ from the total VDoS ${\cal D}(\omega)$, which is experimentally accessible through various scattering techniques, the $\omega^4$ tail of ${\cal D}_{\rm G}(\omega)$ lacked direct experimental support. We develop a procedure to extract ${\cal D}_{\rm G}(\omega)$ from the measured ${\cal D}(\omega)$, based on recent advances in understanding low-frequency excitations in glasses, and apply it to available datasets for diverse glasses. The resulting analysis indicates that the $\omega^4$ tail of the nonphononic vibrational spectra of glasses is nontrivially consistent with a broad range of experimental observations. It also further supports that ${\cal D}_{\rm G}(\omega)$ makes an additive contribution to ${\cal D}(\omega)$.
\end{abstract}

\maketitle

\section{Introduction}\vspace{-0.2cm}

Glasses, which are nonequilibrium disordered materials, reveal properties that are distinct from their crystalline counterparts~\cite{phillips1972tunneling,anderson1972anomalous,soft_potential_model_1991,pohl_review,ramos2022book}. Many of these properties are related to the low-frequency part of the material's vibrational spectra. It is now well established that while both glassy and crystalline materials feature low-frequency phonons, which are related to global continuous symmetries independently of the underlying material structure~\cite{chaikin_lubensky_book}, glasses also host low-frequency, nonphononic excitations~\cite{soft_potential_model_1987,Gurevich2003,Schirmacher_prl_2007,tanaka_boson_peak_2008,eric_boson_peak_emt,JCP_Perspective}. Extensive recent studies of computer glasses, demonstrated that these low-frequency, nonphononic excitations are quasilocalized in space~\cite{JCP_Perspective}, as opposed to the extended nature of low-frequency phonons. Moreover, it has been shown that the quasilocalized, nonphononic excitations follow a non-Debye VDoS that grows as ${\cal D}_{\rm G}(\omega)\!\sim\!\omega^4$ at low vibrational frequencies $\omega$, independently of composition, interatomic interaction, dimensionality and the nonequilibrium history of a glass~\cite{JCP_Perspective,modes_prl_2016,ikeda_pnas,modes_prl_2018,LB_modes_2019,pinching_pnas,modes_prl_2020,universal_VDoS_ip,2d_spectra_jcp_2022,atsushi_2d_pinning_jcp_2023}.

Quasilocalized, nonphononic excitations generically coexist and hybridize with extended phonons~\cite{phonon_widths,boson_peak_2d_jcp_2023}, which follow Debye's VDoS ${\cal D}_{\rm D}(\omega)\!\sim\!\omega^{\dbar-1}$ (in $\dbar$ dimensions), such that it is difficult to determine their spatial structure and number. The above-mentioned progress in computer glasses has been enabled by the development of tools to disentangle the two species of low-frequency excitations, involving thoughtfully selected finite glass sizes~\cite{modes_prl_2016} and/or the accessibility of atomic-scale quantities~\cite{SciPost2016,episode_1_2020,pseudo_harmonic_prl,david_detecting_qles_pre_2023}. These powerful tools are usually not applicable to laboratory glasses, which feature macroscopic sizes (realizing the thermodynamic limit) and where atomic-scale quantities are largely inaccessible.

Experiments on laboratory glasses aim at probing the total VDoS ${\cal D}(\omega)$ through various scattering techniques (down to a technique-specific frequency), as well as macroscopic response quantities such as the specific heat and elastic wave-speeds. So far, such measurements provided only indirect evidence for the $\omega^4$ tail of the nonphononic vibrational spectra of glasses, mainly through a $T^5$ contribution to the specific heat at low temperatures $T$~\cite{soft_potential_model_1991,ramos_2004}. Recent analysis indicated that while quasilocalized, nonphononic excitations generically hybridize with phonons in space, their number per frequency $\omega$ contributes additively to ${\cal D}(\omega)$~\cite{boson_peak_2d_jcp_2023,additive_structure_2024_JCP}, in line with earlier suggestions~\cite{YANNOPOULOS20064541,KALAMPOUNIAS20064619,Schirmacher_prl_2007,grzegorz_2d_modes_prl_2021}. This physical situation implies that the hybridization of excitations in space occurs over a narrow spectral interval around any vibrational frequency $\omega$ such that their numbers make additive contributions to the VDoS per frequency $\omega$, despite the spatial hybridization.

In three dimensions ($\dbar\=3$), the additive structure ${\cal D}(\omega)\simeq{\cal D}_{\rm D}(\omega)+{\cal D}_{\rm G}(\omega)$ takes the form
\begin{equation}
\label{eq:additive}
{\cal D}(\omega)\simeq A_{\rm D}\,\omega^2 + A_{\rm g}\,\omega^4
\end{equation}
at low frequencies $\omega$, where $A_{\rm D}$ is Debye's prefactor of the phononic contribution (that can be obtained through the elastic wave-speeds~\cite{footnote}) and $A_{\rm g}$ is its analog for the nonphononic $\omega^4$ contribution~\cite{pinching_pnas}. The validity of the additive structure of the 3D VDoS has been very recently established, directly and explicitly, in computer glasses~\cite{additive_structure_2024_JCP}. Consequently, the findings of~\cite{additive_structure_2024_JCP} set the ground for applying Eq.~\eqref{eq:additive} to experimental spectra of glasses, with the aim of disentangling the nonphononic $\omega^4$ contribution. Finally, the very recent application of Eq.~\eqref{eq:additive} to experimental ${\cal D}(\omega)$ and $A_{\rm D}$ data of boron oxide glasses provided preliminary evidence for the $\omega^4$ tail of ${\cal D}_{\rm G}(\omega)$~\cite{moriel2024boson}.

Here, we employ the mathematical properties of Eq.~\eqref{eq:additive} to develop a procedure to extract the $\omega^4$ tail of ${\cal D}_{\rm G}(\omega)$ from the experimentally measured ${\cal D}(\omega)$. ${\cal D}(\omega)$, which corresponds to harmonic, zero-temperature vibrational modes (excitations), is commonly probed by scattering techniques, mainly neutron and x-ray scattering~\cite{buchenau1999neutron,taraskin1997connection}. The latter inevitably involve finite temperatures, which potentially imply the intervention of some anharmonicity and relaxational processes~\cite{buchenau1999neutron}, and also involve various analysis transformations~\cite{taraskin1997connection}. Yet, it is commonly concluded, e.g., in~\cite{buchenau1999neutron,taraskin1997connection}, that these measurements performed at relatively low temperatures are reliable and indeed the resulting ${\cal D}(\omega)$ is extensively reported in the literature.

We apply the developed procedure to a variety of VDoS ${\cal D}(\omega)$ available in the literature for diverse glassy materials. The results provide experimental evidence for the $\omega^4$ tail of the nonphononic VDoS ${\cal D}_{\rm G}(\omega)$ in glasses, and support the additive structure of ${\cal D}(\omega)$ in Eq.~\eqref{eq:additive}. The latter indicates that the low-frequency vibrational spectra of glasses are exclusively populated by hybridized phonons and quasilocalized, nonphononic excitations.

\section{Basic relations and analysis methodology}\vspace{-0.2cm}

Equation~\eqref{eq:additive} implies the following relations
\begin{eqnarray}
\label{eq:w4}
{\cal D}(\omega)-A_{\rm D}\,\omega^2 &\xrightarrow[\omega\,\to\,0]{}& A_{\rm g}\,\omega^4 \ ,\\
\label{eq:A_D}
{\cal D}(\omega)/\omega^2 &\xrightarrow[\omega\,\to\,0]{}& A_{\rm D} \ ,
\end{eqnarray}
in the limit of low frequencies, $\omega\!\to\!0$, and ${\cal D}(\omega)/\omega^2$ is known as the reduced VDoS. We use the term `Debye's plateau' to refer to the low-frequency plateau (constant) of the reduced VDoS, which also corresponds to Debye's prefactor $A_{\rm D}$. We distinguish between the two because they are independently measurable, and may differ due to inaccuracies in the measurements, see below.

For the relations in Eqs.~(\ref{eq:w4})-(\ref{eq:A_D}) to be realized in an experimentally measured VDoS ${\cal D}(\omega)$, the latter should go down to sufficiently low frequencies $\omega$. In order to establish criteria for a given ${\cal D}(\omega)$ to qualify for the analysis to be performed below, we first note that the nonphononic VDoS ${\cal D}_{\rm G}(\omega)\={\cal D}(\omega)-A_{\rm D}\,\omega^2$ has been very recently shown to feature an intrinsic peak/maximum, at a frequency denoted as $\omega_{\rm p}$~\cite{moriel2024boson}. The latter should be distinguished from the peak/maximum of the reduced VDoS ${\cal D}(\omega)/\omega^2$, known as the conventional boson peak occurring at $\omega_{_{\rm BP}}$, as extensively discussed in~\cite{moriel2024boson}. For Eq.~\eqref{eq:w4} to be revealed in the experimental data, i.e., for the $\omega^4$ power-law tail to be observed, the measurement of ${\cal D}(\omega)$ should go down to frequencies well below $\omega_{\rm p}$, which is our first qualification criterion.

For the relation in Eq.~\eqref{eq:A_D} to be experimentally realized, the reduced VDoS should reveal a Debye's plateau. A slightly weaker condition would be that the reduced VDoS reveals significant curvature that reasonably indicates a convergence towards a plateau, if lower frequencies would have been probed. These constitute our second qualification criterion. Suppose now that a given experimentally measured ${\cal D}(\omega)$ is judged to be qualified for the analysis, and in particular that there exists a discrete set of $M$ frequencies $\{\omega_i\}_{i=1}^M$ (with the accompanying $\{{\cal D}(\omega_i)\}_{i=1}^M$) that satisfy the criteria stated above. Our next goal is to develop a procedure that allows to test for the consistency of the $\omega^4$ tail of ${\cal D}_{\rm G}(\omega)$ with the experimental data, in view of Eqs.~(\ref{eq:w4})-(\ref{eq:A_D}).

To that aim, we define $\Delta_i(A_{\rm D})\!\equiv\!(\mathcal{D}(\omega_i)\!-\!A_{\rm D}\,\omega_i^2)/\omega_i^4$ based on $\{\omega_i\}_{i=1}^M$, where $A_{\rm D}$ is an unknown parameter. If the $\omega^4$ tail is consistent with experimental data, then one expects that the correct value of Debye's prefactor $A_{\rm D}$ would be such that $\{\Delta_i\}_{i=1}^M$ fluctuate around a constant. Here, the constant corresponds to $A_{\rm g}$, cf.~Eq.~\eqref{eq:w4}, and the fluctuations correspond to experimental measurement noise. Consequently, we consider the coefficient of variation $\eta_{_4}(A_{\rm D})\!\equiv\!\hbox{std}[\{\Delta_i(A_{\rm D)}\}_{i=1}^M]/|\hbox{mean}[\{\Delta_i(A_{\rm D})\}_{i=1}^M]|$, which is a dimensionless measure of the agreement of an experimental dataset with the $\omega^4$ tail, and select $A_{\rm D}$ such that it minimizes $\eta_{_4}(A_{\rm D})$, serving as our objective function~\cite{footnote_Matlab}.

Once $A_{\rm D}$ is selected, $A_{\rm g}\=\hbox{mean}[\{\Delta_i(A_{\rm D})\}_{i=1}^M]$ automatically follows. In order to assess the agreement between the outcome of this single-parameter procedure and the prediction in Eq.~\eqref{eq:additive} (and hence in Eqs.~(\ref{eq:w4})-(\ref{eq:A_D})), we invoke 3 physical criteria to be met: (i) The minimal value of $\eta_{_4}$ should be much smaller than unity. (ii) The fluctuations around $A_{\rm g}$ should be qualified as reasonable experimental measurement noise (e.g., not reveal systematic functional trends and/or clear biases). (iii) The selected $A_{\rm D}$ value should agree with Debye's plateau according to Eq.~\eqref{eq:A_D}.

These assessment criteria pose rather strong self-consistency constraints on the developed, single-parameter procedure. These would provide us with quantitative tools to assess the consistency of available experimental datasets with the prediction in Eq.~\eqref{eq:additive}, even in view of the rather limited experimental scaling regime implied by the smallest vibrational frequencies probed by current scattering techniques. In addition, in situations in which $A_{\rm D}$ is also independently extracted in the experiments based on elastic wave-speeds measurements, we compare it to the $A_{\rm D}$ selected in the analysis.

A complementary analysis to the one described above can be obtained from Eq.~\eqref{eq:additive} by dividing it by $\omega^2$ and plotting ${\cal D}(\omega)/\omega^2$ against $\omega^2$. The latter is predicted to follow a linear function, with slope $A_{\rm g}$ and intercept $A_{\rm D}$, in the low-frequency regime (yet, over a finite range of frequencies $\omega$, not just the strict $\omega\!\to\!0$ limit as in Eq.~\eqref{eq:A_D}). The main merit of this complementary approach is that the prediction of a linear variation of ${\cal D}(\omega)/\omega^2$ as a function of $\omega^2$, i.e., ${\cal D}(\omega)/\omega^2\=A_{\rm D}+A_{\rm g}\,\omega^2$, is parameter-free. We employ this complementary analysis to further test the validity of Eq.~\eqref{eq:additive}. Finally, we also construct a dimensionless ratio out of the extracted $A_{\rm D}$ and $A_{\rm g}$ (see below), and compare it to the corresponding ratio available through computer glass studies, which is yet another self-consistency test.

\begin{table}[ht!]
    \centering
\begin{tabular}{p{4.2cm} l}
     \hline
     \hline
     Glassy material & Data source \\
     \hline
     ~Toluene & ~Fig.~2 (22 K dataset)~\cite{chumakov2004collective} \\
     ~Dibutyl phthalate & ~Fig.~2 (22 K dataset)~\cite{chumakov2004collective} \\
     ~Polybutadiene & ~Fig.~12 (60 K dataset)~\cite{zorn1995neutron} \\
     ~Glycerol & ~Fig.~4 (80 K dataset)~\cite{wuttke1995fast} \\
     ~Ambient silica & ~Fig.~1c (solid circles)~\cite{chumakov2014role} \\
     ~Densified silica & ~Fig.~1d (solid circles)~\cite{chumakov2014role} \\
     \hline
     \hline
\end{tabular}
\caption{The glassy materials for which the VDoS ${\cal D}(\omega)$ has been analyzed and the precise experimental source from which each dataset was digitized.}
\label{tab:sources}
\end{table}

\section{Analysis of experimental data}\vspace{-0.2cm}

We apply the above procedure to 6 experimentally measured VDoS ${\cal D}(\omega)$ that were judged to qualify for the analysis according to the above stated qualification criteria. These correspond to diverse glassy materials, of different compositions and nonequilibrium histories, as detailed in Table~\ref{tab:sources} and in the references therein. The experimental data were obtained using various scattering techniques (e.g., nuclear inelastic scattering, neutron scattering and inelastic x-ray scattering), as detailed in the references provided in Table~\ref{tab:sources}.

Each dataset was obtained by digitizing the published experimental data (the precise figure sources are detailed in Table~\ref{tab:sources}) using a commercial software~\cite{WebPlotDig}. Whenever data were presented for several measurement temperatures, we digitized the lowest temperature one (indicated in Table~\ref{tab:sources}) in order to minimize anharmonic effects~\cite{buchenau1999neutron}. Whenever $A_{\rm D}$ was independently measured and reported, it was extracted as well.
\begin{figure}[ht!]
\center
\includegraphics[width=0.49\textwidth]{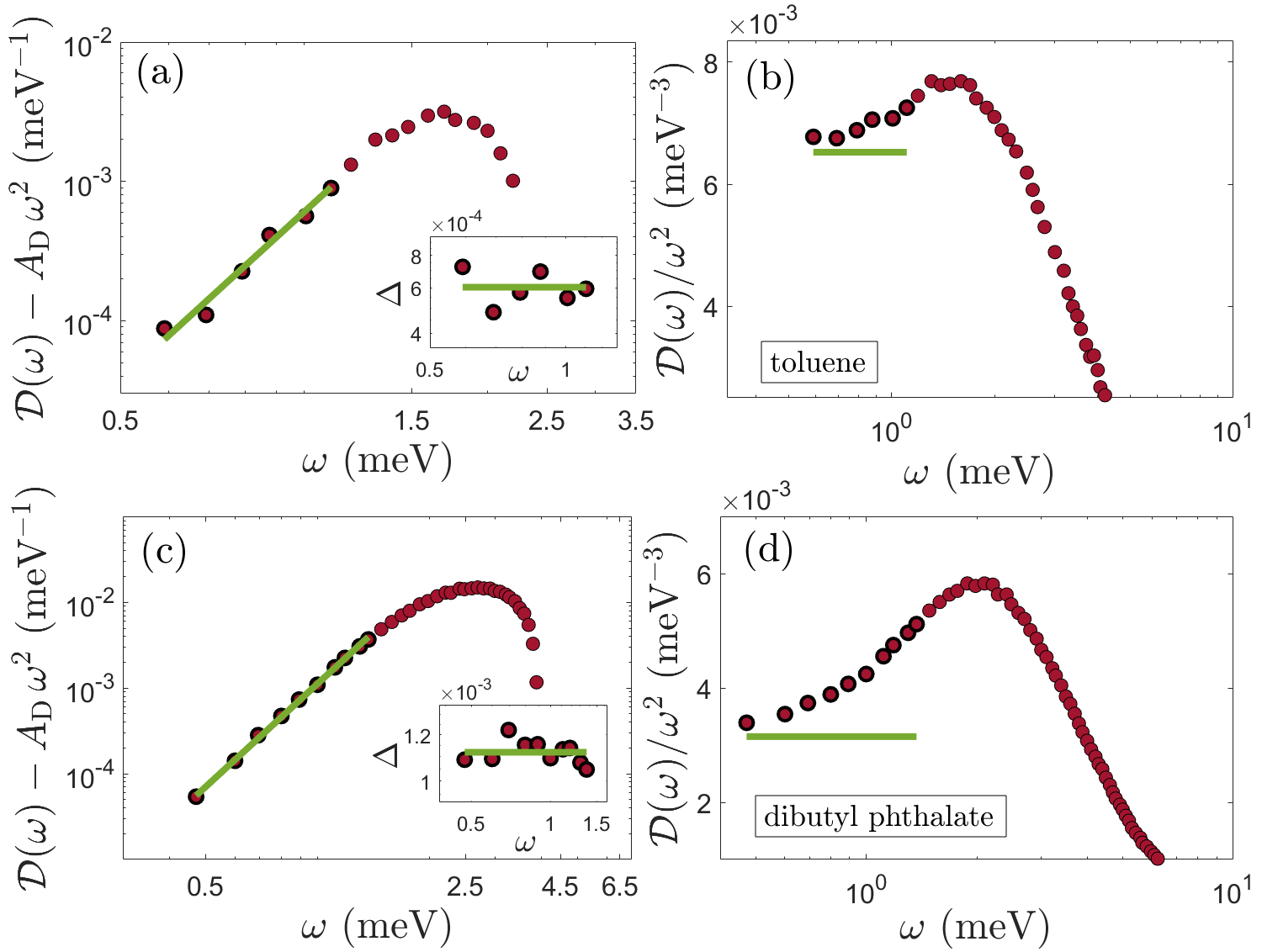}
\caption{(top row) Analysis of ${\cal D}(\omega)$ of glassy toluene, see Table~\ref{tab:sources}.  (a) ${\cal D}(\omega)-A_{\rm D}\,\omega^2$, where $A_{\rm D}\!\simeq\!6.5\times 10^{-3}\,\text{meV}^{-3}$ was determined by minimizing $\eta_{_4}$, see text. $M\!=\!6$ data points were analyzed (thick-boundary circles), where the ratio of the largest frequency analyzed, $\omega_{_{\rm M}}$, to the intrinsic boson peak~\cite{moriel2024boson}, $\omega_{\rm p}$, is $\omega_{_{\rm M}}/\omega_{\rm p}\!\simeq\!0.65$. The green line corresponds to $A_{\rm g}\,\omega^4$, with $A_{\rm g}\!\simeq\! 6.0\times 10^{-4}\,\text{meV}^{-5}$ (automatically emerging from the analysis, see text), corresponding to a minimal value of $\eta_{_{_4}}\!\simeq\!0.15$. The corresponding  dimensionless frequency ratio is $A_{\rm g}^{-1/5}/\omega_{_{\rm D}}\!\simeq\!0.57$ (see text). (inset) $\Delta(\omega)$ of the analyzed data points, where the horizontal green line corresponds to $A_{\rm g}$ (see text). The units are not indicated (these are $\text{meV}^{-5}$ for $\Delta$ and meV for $\omega$). (b) ${\cal D}(\omega)/\omega^2$, where the horizontal green line corresponds to $A_{\rm D}$ of panel (a).
(bottom row) Analysis of ${\cal D}(\omega)$ of glassy dibutyl phthalate, see Table~\ref{tab:sources}.
(c) Same as panel (a), with $A_{\rm D}\!\simeq\!3.1\times 10^{-3}\,\text{meV}^{-3}$, $M\!=\!10$, $\omega_{_{\rm M}}/\omega_{\rm p}\!\simeq\!0.51$, $A_{\rm g}\!\simeq\!1.1\times 10^{-3}\,\text{meV}^{-5}$, $\eta_{_{4}}\!\simeq\!0.05$ and $A_{\rm g}^{-1/5}/\omega_{\rm D}\!\simeq\!0.40$. (d) Same as panel (b).}
\label{fig:fig1}
\end{figure}

In Figs.~\ref{fig:fig1}a-b, we present the analysis of ${\cal D}(\omega)$ of glassy toluene (see Table~\ref{tab:sources}), which is the first experimental dataset discussed. It is also used to introduce the presentation format of our results and the application of the above stated assessment criteria. In Fig.~\ref{fig:fig1}a, we plot on a double-logarithmic scale ${\cal D}_{\rm G}(\omega)\={\cal D}(\omega)\!-\!A_{\rm D}\,\omega^2$ (brown circles) using $A_{\rm D}$ that minimizes $\eta_{_4}(A_{\rm D})$ (see caption). All data points used in the analysis (marked by the thick-boundary circles) have frequencies below the peak $\omega_{\rm p}$ (see caption), as required by our first qualification criterion. The superimposed solid green line corresponds to $A_{\rm g}\,\omega^4$, with $A_{\rm g}\=\hbox{mean}[\{\Delta_i(A_{\rm D})\}_{i=1}^M]$ (see caption). It is observed that the obtained $A_{\rm g}\,\omega^4$ is in very good quantitative agreement with the experimental data, as is also clear from the minimal value of the objective function, $\eta_{_4}\!\simeq\!0.15\!\ll\!1$.

In the inset of Fig.~\ref{fig:fig1}a, we present $\Delta(\omega)$, which appears to reveal a legitimate $\omega$-dependent noise. Finally, in Fig.~\ref{fig:fig1}b, we plot the reduced VDoS ${\cal D}(\omega)/\omega^2$ on a semi-logarithmic scale, marking by thick-boundary circles the very same data points used in the analysis in panel (a), similarly marked therein. The data indeed approach a Debye's plateau at the lowest frequencies, as required by our second qualification criterion. We then superimpose in the solid green line the value of $A_{\rm D}$ obtained in panel (a), which is observed to remarkably agree with Debye's plateau. Consequently, the glassy toluene analysis in Figs.~\ref{fig:fig1}a-b is nontrivially consistent with the $\omega^4$ tail of the nonphononic VDoS ${\cal D}_{\rm G}(\omega)$ of glasses.

In Figs.~\ref{fig:fig1}c-d, the same analysis is performed for glassy dibutyl phthalate (see Table~\ref{tab:sources}). The consistency of the data with the prediction in Eq.~\eqref{eq:additive} is even stronger than that of the toluene data in Figs.~\ref{fig:fig1}a-b, as is visually evident and manifested by the very small value of the objective function, $\eta_{_4}\!\simeq\!0.05\!\ll\!1$. These results provide additional experimental evidence for the $\omega^4$ tail of the nonphononic spectra of glasses. The experiments giving rise to the datasets analyzed in Fig.~\ref{fig:fig1} did not independently measure Debye's prefactor $A_{\rm D}$~\cite{chumakov2004collective}. Next, we analyze two additional datasets, for different glassy materials, in which independent measurements of Debye's prefactor $A_{\rm D}$ are available.

\begin{figure}[ht!]
\center
\includegraphics[width=0.49\textwidth]{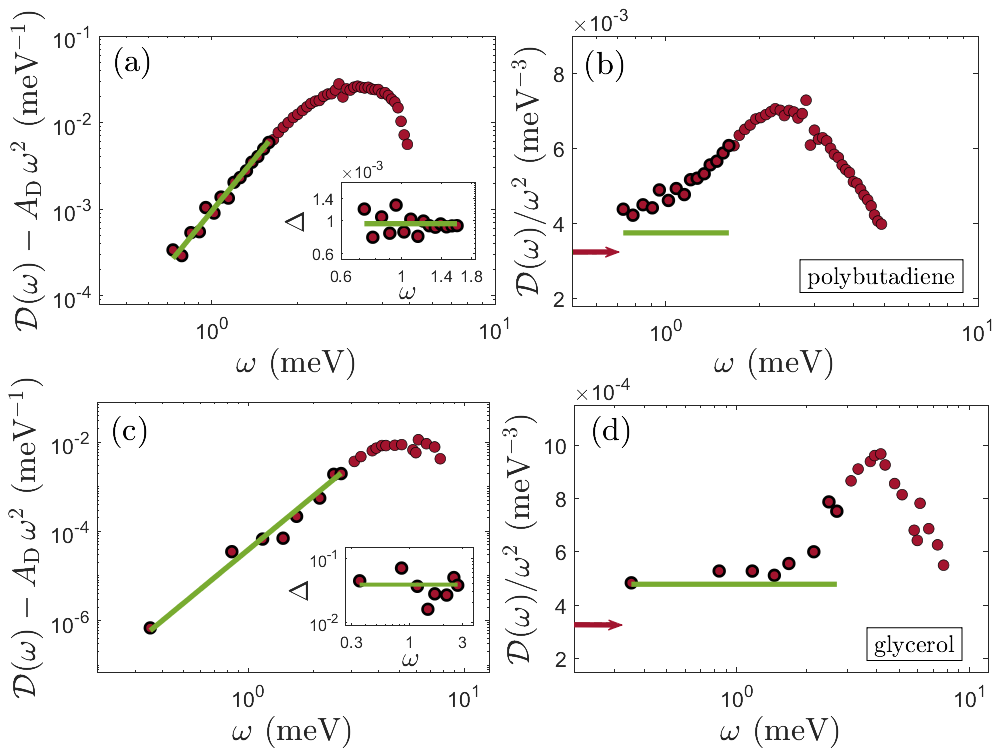}
\caption{(top row) Analysis of ${\cal D}(\omega)$ of glassy polybutadiene, see Table~\ref{tab:sources}. (a) Same format as in Fig.~\ref{fig:fig1}a, with $A_{\rm D}\!\simeq\!3.8\times 10^{-3}\,\text{meV}^{-3}$, $M\!=\!15$, $\omega_{_{\rm M}}/\omega_{\rm p}\!\simeq\!0.57$, $A_{\rm g}\!\simeq\!9.5\times 10^{-4}\,\text{meV}^{-5}$, $\eta_{_{4}}\!\simeq\!0.15$ and $A_{\rm g}^{-1/5}/\omega_{\rm D}\!\simeq\!0.43$. (b) Same format as in Fig.~\ref{fig:fig1}b, where in addition the brown arrow corresponds to the experimentally determined $A_{\rm D}$ (see Table~\ref{tab:sources} for the source). (bottom row) Analysis of ${\cal D}(\omega)$ of glassy glycerol, see Table~\ref{tab:sources}. Same format as above, with $A_{\rm D}\!\simeq\!4.8\times 10^{-4}\,\text{meV}^{-3}$, $M\!=\!8$, $\omega_{_{\rm M}}/\omega_{\rm p}\!\simeq\!0.43$, $A_{\rm g}\!\simeq\!3.9\times 10^{-5}\,\text{meV}^{-5}$, $\eta_{_{4}}\!\simeq\!0.43$ and $A_{\rm g}^{-1/5}/\omega_{\rm D}\!\simeq\!0.41$.}
\label{fig:fig2}
\end{figure}

In the top row of Fig.~\ref{fig:fig2}, we present results for glassy polybutadiene (see Table~\ref{tab:sources}), using exactly the same presentational format as in Fig.~\ref{fig:fig1}. The results in Fig.~\ref{fig:fig2}a reveal excellent agreement between the experimental data and the $\omega^4$ tail for the selected $A_{\rm D}$ (see caption). $A_{\rm D}$ is superimposed (green line) on the reduced VDoS in Fig.~\ref{fig:fig2}b and appears to be consistent with an extrapolated Debye's plateau. The reduced VDoS in Fig.~\ref{fig:fig2}b reveals curvature on the semi-logarithmic scale used, but not yet a plateau. Interestingly, the independently measured $A_{\rm D}$ (brown arrow) is smaller than the extracted $A_{\rm D}$ value and may appear less consistent with an extrapolated Debye's plateau, though a strictly quantitative statement cannot be made in the absence of a clear Debye's plateau.

The situation is different in the bottom row of Fig.~\ref{fig:fig2}, where we present results for glassy glycerol (see Table~\ref{tab:sources}). The results in Fig.~\ref{fig:fig2}c yet again reveal reasonably good agreement between the experimental data and the $\omega^4$ tail for the selected $A_{\rm D}$, see figure caption for the extracted parameters. Note that the value of the minimal objective function is smaller than unity, $\eta_{_4}\!\simeq\!0.43$, yet it is larger than the corresponding values in the analyses in Fig.~\ref{fig:fig1} (as is also evident from the observed fluctuations). The selected $A_{\rm D}$ is superimposed (green line) on the reduced VDoS in Fig.~\ref{fig:fig2}d and is perfectly consistent with the observed Debye's plateau. In contrast, the independently measured $A_{\rm D}$ (brown arrow) deviates quite significantly from Debye's plateau, a point that is unfortunately not discussed in the original experimental report~\cite{wuttke1995fast}. Note, however, that the elastic measurements used to independently estimate $A_{\rm D}$ in~\cite{wuttke1995fast} were not performed on the same samples (but were rather extracted from Refs.~[22, 23] therein), which in view of the nonequilibrium history dependence of glasses, may explain the deviation.
\begin{figure}[ht!]
\center
\includegraphics[width=0.49\textwidth]{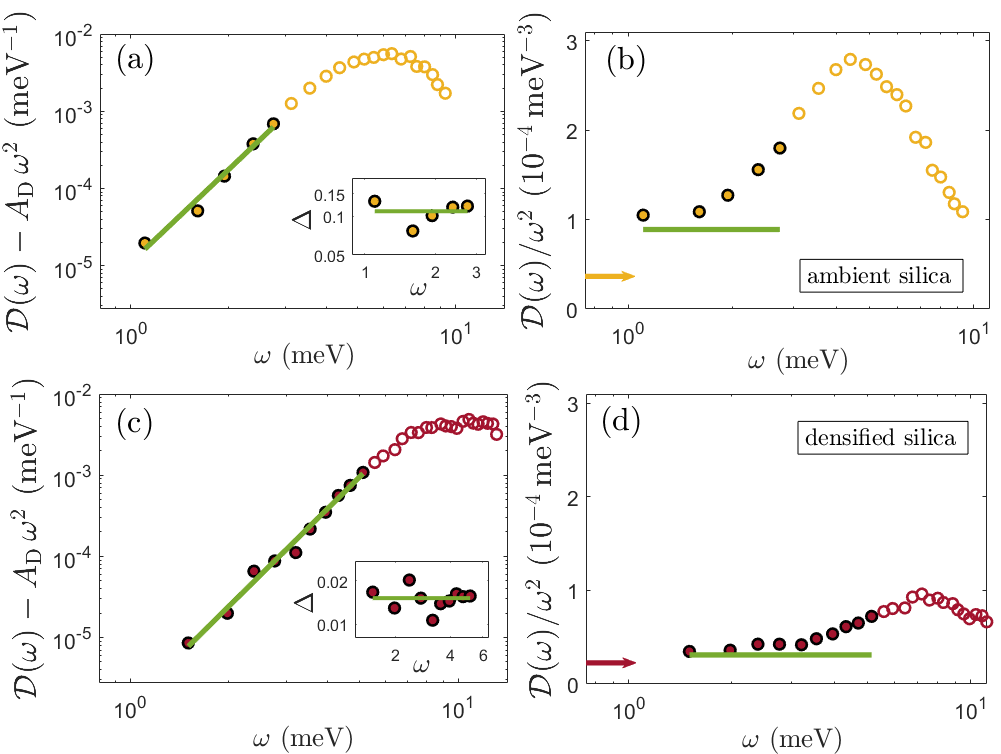}
\caption{(top row) Analysis of ${\cal D}(\omega)$ of silica glass generated under ambient conditions (orange circles), see Table~\ref{tab:sources}. Same format as in Fig.~\ref{fig:fig2}, with $A_{\rm D}\!\simeq\!8.9\times 10^{-5}\,\text{meV}^{-3}$, $M\!=\!5$, $\omega_{_{\rm M}}/\omega_{\rm p}\!\simeq\!0.43$, $A_{\rm g}\!\simeq\!1.1\times 10^{-5}\,\text{meV}^{-5}$, $\eta_{_{4}}\!=\!0.19$ and $A_{\rm g}^{-1/5}\!/\omega_{\rm D}\!\simeq\!0.30$. (bottom row) Analysis of ${\cal D}(\omega)$ of densified silica glass (brown circles), see Table~\ref{tab:sources}. Same format as in Fig.~\ref{fig:fig2}, with $A_{\rm D}\!\simeq\!3.1\times 10^{-5}\,\text{meV}^{-3}$, $M\!=\!10$, $\omega_{_{\rm M}}/\omega_{\rm p}\!\simeq\!0.48$, $A_{\rm g}\!\simeq\!0.2\times 10^{-5}\,\text{meV}^{-5}$, $\eta_{_{4}}\!\simeq\!0.17$ and $A_{\rm g}^{-1/5}\!/\omega_{\rm D}\!\simeq\!0.32$. Note that a different color scheme is used to highlight that here we consider the same glass composition of different nonequilibrium histories (see text).}
\label{fig:fig3}
\end{figure}

In Figs.~\ref{fig:fig1}-\ref{fig:fig2}, we analyzed experimental ${\cal D}(\omega)$ data for 4 glassy materials, differing in composition, interatomic interactions and possibly preparation protocols. In Fig.~\ref{fig:fig3}, we turn our attention to a pair of glasses that share the same composition (and hence interatomic interactions), but differ in their nonequilibrium history. Specifically, we consider silica glasses at different densities, which opens the way for comparing different glassy states on equal footing, where our a priori expectation is that both $A_{\rm g}$ and $A_{\rm D}$ decrease upon densification.

In the top row of Fig.~\ref{fig:fig3}, we present the analysis of ${\cal D}(\omega)$ of silica glass formed under ambient conditions (orange circles), see Table~\ref{tab:sources}. The results reveal excellent agreement with the $\omega^4$ tail of the nonphononic VDoS ${\cal D}_{\rm G}(\omega)$ (panel (a)) and the selected $A_{\rm D}$ is in great quantitative agreement with the respective Debye's plateaus (panel (b)), while the independently measured $A_{\rm D}$ (orange arrow) quite significantly deviates from the respective Debye's plateau. The authors of~\cite{chumakov2014role} speculate that the observed deviation might be due to some not accounted for effects, e.g., mechanical instability or anharmonicity, see footnote [40] therein.

In the bottom row of Fig.~\ref{fig:fig3}, we present the corresponding analysis for densified silica glass (brown circles), which also reveals strong consistency with Eq.~\eqref{eq:additive}, in terms of all of the considered assessment criteria. Moreover, the obtained $A_{\rm g}$ and $A_{\rm D}$ in both rows of the figure (visually observed in the figure and quantitatively reported in the caption) indeed decrease with densification, as expected. Overall, the results in Fig.~\ref{fig:fig3} lend additional support to our main finding.
\begin{figure*}[ht!]
\center
\includegraphics[width=\textwidth]{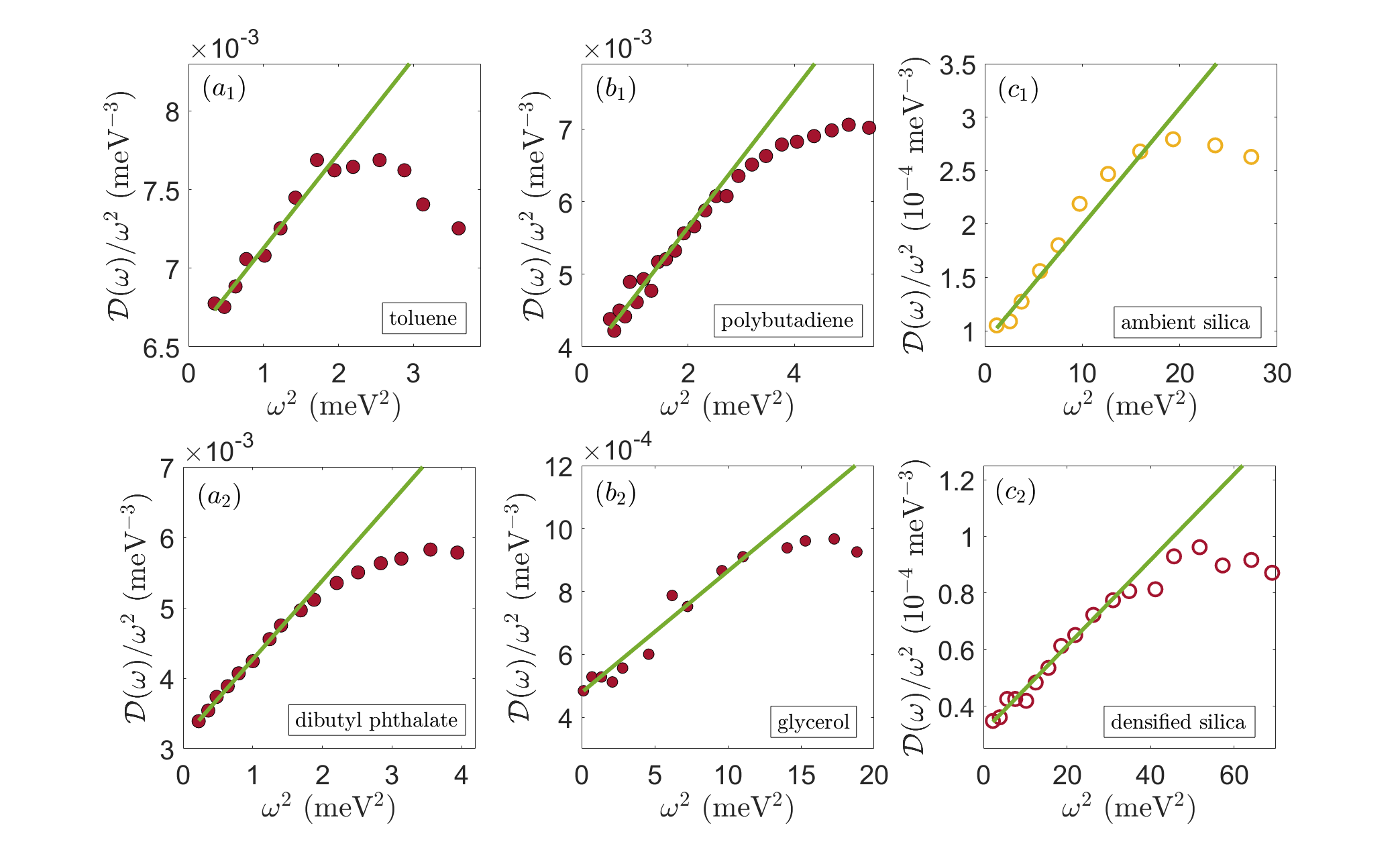}
\vspace{-0.25cm}
\caption{A complementary analysis to the one presented in Figs.~\ref{fig:fig1}-\ref{fig:fig3}, where the same ${\cal D}(\omega)$ datasets as presented therein are used (same symbol styles and colors), but this time replotted as ${\cal D}(\omega)/\omega^2$ vs.~$\omega^2$. Equation~\eqref{eq:additive} predicts that in this representation the low-frequency tail follows a linear function, with slope $A_{\rm g}$ and intercept $A_{\rm D}$. The different panels show ${\cal D}(\omega)/\omega^2$ vs.~$\omega^2$ for (a$_1$) the glassy toluene data presented in Fig.~\ref{fig:fig1}b, (a$_2$) the glassy dibutyl phthalate data presented in Fig.~\ref{fig:fig1}d, (b$_1$) the glassy polybutadiene data presented in Fig.~\ref{fig:fig2}b, (b$_2$) the glassy glycerol data presented in Fig.~\ref{fig:fig2}d, (c$_1$) the ambient silica data presented in Fig.~\ref{fig:fig3}b and (c$_2$) the densified silica data presented in Fig.~\ref{fig:fig3}d. The superimposed green lines in all panels are not linear fits, but rather lines with slope $A_{\rm g}$ and intercept $A_{\rm D}$, where the values of $A_{\rm g}$ and $A_{\rm D}$ in each panel are those extracted in the corresponding analysis in Figs.~\ref{fig:fig1}-\ref{fig:fig3}.}
\label{fig:fig4}
\end{figure*}

To further support our main finding, we plot in Fig.~\ref{fig:fig4} ${\cal D}(\omega)/\omega^2$ as a function of $\omega^2$ for the 6 datasets presented in Figs.~\ref{fig:fig1}-\ref{fig:fig3}. As explained above, Eq.~\eqref{eq:additive} predicts that this function is linear at small $\omega^2$, taking the form ${\cal D}(\omega)/\omega^2\=A_{\rm D}+A_{\rm g}\,\omega^2$, and this functional form prediction is parameter-free. The results presented in Fig.~\ref{fig:fig4} strongly support this prediction for all 6 datasets considered, in agreement with the analysis presented in Figs.~\ref{fig:fig1}-\ref{fig:fig3}. It is important to note that we did not extract the slope $A_{\rm g}$ and intercept $A_{\rm D}$ using this complementary analysis, which would have involved a two-parameter linear fit, but rather used the values of $A_{\rm D}$ and $A_{\rm g}$ extracted in the analysis presented in Figs.~\ref{fig:fig1}-\ref{fig:fig3} (based on a single-parameter procedure) to superimpose the green lines in the 6 panels of Fig.~\ref{fig:fig4}. The resulting quantitative agreement with the experimental data at small $\omega^2$ is very good.

The analyses presented above tested the consistency of the ${\cal D}_{\rm G}(\omega)\!\sim\!\omega^4$ tail scaling with experimental data for various glasses. In view of the frequency range of scaling accessible through current experimental techniques, below $\omega_{\rm p}$, it is legitimate to ask to what extent the data considered might be consistent with ${\cal D}_{\rm G}(\omega)\!\sim\!\omega^\beta$, with $\beta\!\ne\!4$. In~\cite{SM}, we present analyses of the dibutyl phthalate dataset (smallest $\eta_{_4}$, $\eta_{_4}\!\simeq\!0.05$, see Fig.~\ref{fig:fig1}c-d) and the glycerol dataset (largest $\eta_{_4}$, $\eta_{_4}\!\simeq\!0.43$, see Fig.~\ref{fig:fig2}c-d) with $\beta\=3.5$ and $\beta\=4.5$. The results presented in~\cite{SM}, and their discussion in view of our assessment criteria, indicate that $\beta\!\approx\!4$ is rather robustly selected by the experimental data, with an uncertainty window that shrinks with $\eta_{_4}$.

Our analysis gives rise to the VDoS prefactors $A_{\rm D}$ and $A_{\rm g}$, which are associated with characteristic vibrational frequencies of glasses. Specifically, $A_{\rm g}^{-1/5}$ is a characteristic frequency scale of quasilocalized, nonphononic excitations, while Debye's frequency $\omega_{_{\rm D}}\=(3/A_{\rm D})^{1/3}$ is a characteristic frequency scale of phononic excitations~\cite{chaikin_lubensky_book,footnote}. The dimensionless frequency ratio $A_{\rm g}^{-1/5}\!/\omega_{_{\rm D}}$ has been measured in a variety of computer glasses. In~\cite{SM}, we collect available data on various computer glass models under a wide variety of nonequlibrium histories. We show that $A_{\rm g}^{-1/5}\!/\omega_{_{\rm D}}$ spans the range $0.22\!-\!0.73$, including deeply supercooled glassy states, expected to be relevant for laboratory glasses.

In the captions of Figs.~\ref{fig:fig1}-\ref{fig:fig3}, we report the values of $A_{\rm g}^{-1/5}\!/\omega_{_{\rm D}}$ emerging from our analysis of experimental data for the 6 glassy materials considered, found to span the range $0.30\!-\!0.57$. The latter quite remarkably overlaps the corresponding range for computer glasses (towards its lower part), lending additional support to the experimental evidence provided in this work for the $\omega^4$ tail of the nonphononic vibrational spectra of glasses.

\section{Summary and outlook}\vspace{-0.2cm}

We developed a procedure to compare the predictions of Eq.~\eqref{eq:additive} to experimental data for various glassy materials. Equation~(\ref{eq:additive}) predicts that (i) the low-frequency regime of the total VDoS ${\cal D}(\omega)$ is exclusively populated by quasilocalized nonphononic vibrations and phonons, (ii) the VDoSs of the two species contribute additively to ${\cal D}(\omega)$, (iii) the tail of the nonphononic VDoS scales as ${\cal D}_{\rm G}(\omega)\!\sim\!\omega^4$. The analysis indicates that Eq.~\eqref{eq:additive} is nontrivially consistent with a broad range of experimental data, and hence provides experimental evidence for its validity, and for the nonphononic $\omega^4$ tail in particular. We stress that while the frequency range over which this scaling is observed is limited by experimental capabilities, ${\cal D}_{\rm G}(\omega)$ varies substantially, e.g., by more than 3 orders of magnitude in Fig.~\ref{fig:fig2}c, due to the strong $\sim\!\omega^4$ variation.

These findings are of particular importance as low-frequency nonphononic excitations in glasses crucially affect sound attenuation, the specific heat, thermal transport and plastic deformation (see, for example,~\cite{ramos2022book,JCP_Perspective}), and are also known to constitute the elementary building blocks of boson peak excitations~\cite{boson_peak_2d_jcp_2023,moriel2024boson}.

The simple procedure can be straightforwardly applied to other experimental ${\cal D}(\omega)$ datasets of glasses. Future experimental techniques will hopefully be able to probe smaller frequencies, further constraining the emerging physical picture. Future work should also go beyond the tail regime and consider the boson peak regime, in particular the intrinsic peak $\omega_{\rm p}$ of ${\cal D}_{\rm G}(\omega)$ (observed on the left panels of each figure above) and the conventional peak $\omega_{_{\rm BP}}$ of the reduced VDoS ${\cal D}(\omega)/\omega^2$ (observed on the right panels of each figure above), following recent developments~\cite{moriel2024boson}.

{\em Acknowledgements}. This work has been supported by the Israel Science Foundation (ISF grant no.~403/24). A.M.~acknowledges support from the James S.~McDonnell Foundation Postdoctoral Fellowship Award in Complex Systems (\url{https://doi.org/10.37717/2021-3362}). E.B.~acknowledges support from the Ben May Center for Chemical Theory and Computation and the Harold Perlman Family.
\vspace{0.3cm}

\hspace{-0.1cm}{\bf Supplementary Material}\\

The Supplementary Material file is appended below.

\clearpage

\onecolumngrid
\begin{center}
              \textbf{\Large Supplemental material}
\end{center}

\setcounter{equation}{0}
\setcounter{figure}{0}
\setcounter{section}{0}
\setcounter{subsection}{0}
\setcounter{table}{0}
\setcounter{page}{1}
\makeatletter
\renewcommand{\theequation}{S\arabic{equation}}
\renewcommand{\thefigure}{S\arabic{figure}}
\renewcommand{\thesection}{S-\Roman{section}}
\renewcommand{\thesubsection}{S-\Roman{subsection}}
\renewcommand*{\thepage}{S\arabic{page}}
\twocolumngrid

The goal of this document is to discuss the robustness of the experimental evidence for the $\sim\!\omega^\beta$ scaling of the low-frequency tail of the nonphononic spectra of glasses, with $\beta\=4$. The $\beta\=4$ analysis presented in the manuscript is supplemented here in Sect.~\ref{sec:different_beta} with a few examples of analyses involving $\beta\!\ne\!4$, along with an accompanying discussion. Finally, we also provide in Sect.~\ref{sec:frequency_ratio} details about the literature sources of the $A_{\rm g}^{-1/5}\!/\omega_{_{\rm D}}$ ratios of computer glasses, reported in the manuscript.

\section{A\lowercase{nalyses with} $\beta\!\ne\!4$}
\label{sec:different_beta}

We start by generalizing the coefficient of variation defined in the manuscript, which is a central quantity in testing the consistency of Eqs.~(1)-(2) therein with experimental data. Specifically, we define
\begin{equation}
\Delta_i(A_{\rm D};\beta) \equiv \frac{\mathcal{D}(\omega_i)-A_{\rm{D}}\,\omega_i^2}{\omega_i^\beta}\ ,
\label{eq:Delta_beta}
\end{equation}
which conforms with the definition used in the manuscript for $\beta\=4$. Note that the dimension of $\Delta$ varies with $\beta$, and is identical to that of $A_{\rm g}$ (recall that $A_{\rm g}\,\omega^\beta$ is of inverse frequency dimension). Subsequently, we define
\begin{equation}
\eta_{_\beta}(A_{\rm D})\equiv\frac{\hbox{std}[\{\Delta_i(A_{\rm D};\beta)\}_{i=1}^M]}{|\hbox{mean}[\{\Delta_i(A_{\rm D};\beta)\}_{i=1}^M]|} \ ,
\end{equation}
which yet again conforms with the corresponding definition used in the manuscript for $\beta\=4$. We select for the $\beta\!\ne\!4$ analyses the dataset with the smallest $\beta\=4$ minimal (optimal) objective function, i.e., $\eta_{_4}\!\simeq\!0.05$ for glassy dibutyl phthalate (cf.~Fig.~1c in the manuscript), and the one with the largest $\beta\=4$ minimal (optimal) objective function, i.e., $\eta_{_4}\!\simeq\!0.43$ for glassy glycerol (cf.~Fig.~2c in the manuscript). We discuss the results in view of the three assessment criteria formulated in the manuscript.
\begin{figure}[ht!]
\center
\includegraphics[width=0.49\textwidth]{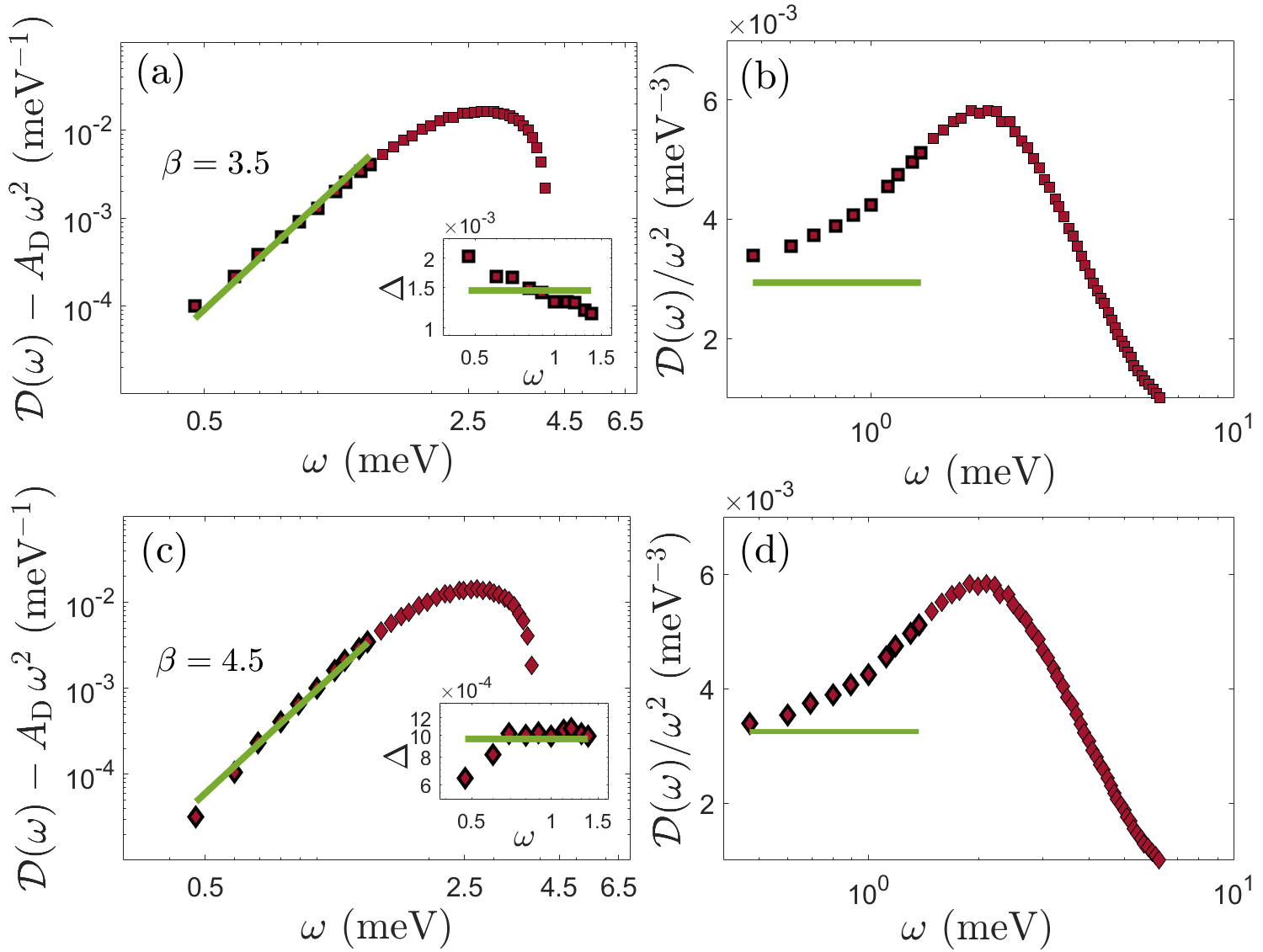}
\caption{Analysis of the same ${\cal D}(\omega)$ of glassy dibutyl phthalate as in Fig.~1c-d in the manuscript (again with $M\!=\!10$ and $\omega_{_{\rm M}}/\omega_{\rm p}\!\simeq\!0.51$), but with $\beta\!\ne\!4$. (top row) $\beta\!=\!3.5$ (brown squares), with $A_{\rm D}\!\simeq\!2.9\times 10^{-3}\,\text{meV}^{-3}$, $A_{\rm g}\!\simeq\!1.5\times 10^{-3}\,\text{meV}^{-4.5}$ and $\eta_{_{3.5}}\!\simeq\!0.03$. (bottom row) $\beta\!=\!4.5$ (brown diamonds), with $A_{\rm D}\!\simeq\!3.2\times 10^{-3}\,\text{meV}^{-3}$, $A_{\rm g}\!\simeq\!9.6\times 10^{-4}\,\text{meV}^{-5.5}$ and $\eta_{_{4.5}}\!\simeq\!0.11$. Note that the green line in panel (a) corresponds to $A_{\rm g}\,\omega^{3.5}$, hence the dimension of $A_{\rm g}$ differs from that of the corresponding quantity for $\beta\!=\!4$ in the manuscript. The same applies to the green line in panel (c), which corresponds to $A_{\rm g}\,\omega^{4.5}$. Note also that we used here different symbols compared to the manuscript figures in order to highlight the different $\beta$ values used.}
\label{fig:figS1}
\end{figure}

In the top row of Fig.~\ref{fig:figS1}, we present the analysis of ${\cal D}(\omega)$ of glassy dibutyl phthalate with $\beta\=3.5$. First, it is observed (Fig.~\ref{fig:figS1}b and its caption) that the selected $A_{\rm D}$ is very similar to the one selected in the manuscript for $\beta\!=\!4$ ($2.9\times 10^{-3}\,\text{meV}^{-3}$ compared to $3.1\times 10^{-3}\,\text{meV}^{-3}$), which is expected since $\beta$ is still significantly larger than 2, hence fluctuations in $\Delta_i(A_{\rm D};\beta)$ are reduced by properly eliminating the dominant $A_{\rm D}\,\omega^2$ contribution to ${\cal D}(\omega)$. Second, it is observed that while the objective function attains a somewhat smaller value of $\eta_{_{3.5}}\!\simeq\!0.03$ (see caption, compared to $\eta_{_{4}}\!\simeq\!0.05$, see caption of Fig.~1 in the manuscript), the inset of Fig.~\ref{fig:figS1}a reveals that the distribution of $\Delta_i$ around its mean appears deterministic (it is a monotonically decreasing function of $\omega$), i.e., it does not appear to correspond to a reasonable experimental measurement noise. This is also visually evident from comparing the $A_{\rm g}\,\omega^{3.5}$ green line to the experimental data in the main panel of Fig.~\ref{fig:figS1}a. Consequently, we conclude that $\beta\=3.5$ is far less consistent with the experimental data for glassy dibutyl phthalate compared to $\beta\=4$.

In the bottom row of Fig.~\ref{fig:figS1}, we present the corresponding analysis for $\beta\=4.5$. First, it appears (Fig.~\ref{fig:figS1}d) that the extrapolated Debye's plateau would intersect the selected $A_{\rm D}$ value. Second, the inset of Fig.~\ref{fig:figS1}c reveals apparently systematic deviations at low frequencies $\omega$, as is also visually observed in the main panel of Fig.~\ref{fig:figS1}c. These deviations give rise to a larger minimal (optimal) value of the objective function, $\eta_{_{4.5}}\!\simeq\!0.11$ (see caption, compared to $\eta_{_{4}}\!\simeq\!0.05$, see caption of Fig.~1 in the manuscript). Consequently, we also conclude that $\beta\=4.5$ is less likely than $\beta\=4$ for the glassy dibutyl phthalate dataset. Taken together, it appears as if the experimental data in this case rather robustly selects $\beta\=4$.
\begin{figure}[ht!]
\center
\includegraphics[width=0.49\textwidth]{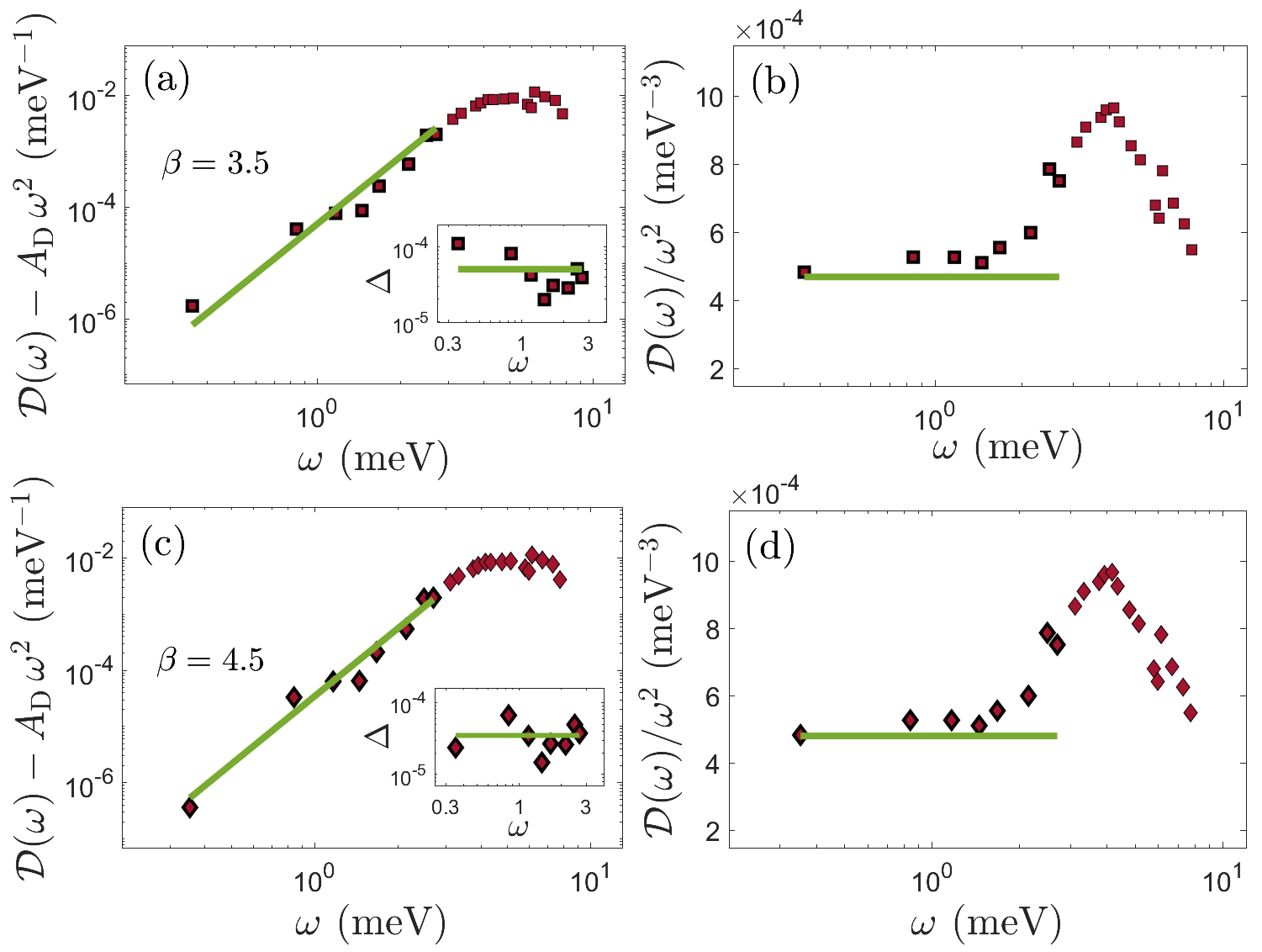}
\caption{Analysis of the same ${\cal D}(\omega)$ of glassy glycerol as in Fig.~2c-d in the manuscript (again with $M\!=\!8$ and $\omega_{_{\rm M}}/\omega_{\rm p}\!\simeq\!0.43$), but with $\beta\!\ne\!4$, in the same format as in Fig.~\ref{fig:figS1}. (top row) $\beta\!=\!3.5$, with $A_{\rm D}\!\simeq\!4.7\times 10^{-4}\,\text{meV}^{-3}$, $A_{\rm g}\!\simeq\!5.1\times 10^{-5}\,\text{meV}^{-4.5}$ and $\eta_{_{3.5}}\!\simeq\!0.36$. (bottom row) $\beta\!=\!4.5$, with $A_{\rm D}\!\simeq\!4.8\times 10^{-4}\,\text{meV}^{-3}$, $A_{\rm g}\!\simeq\!3.5\times 10^{-5}\,\text{meV}^{-5.5}$ and $\eta_{_{4.5}}\!\simeq\!0.61$.}
\label{fig:figS2}
\end{figure}

In the top row of Fig.~\ref{fig:figS2}, we present the analysis of ${\cal D}(\omega)$ of glassy glycerol with $\beta\=3.5$. First, it is observed (Fig.~\ref{fig:figS2}b and the caption) that the selected $A_{\rm D}$ is similar to the one selected in the manuscript for $\beta\=4$ ($4.7\times 10^{-4}\,\text{meV}^{-3}$ compared to $4.8\times 10^{-4}\,\text{meV}^{-3}$). Second, it is observed in the inset of Fig.~\ref{fig:figS2}a that the distribution of $\Delta_i$ around its mean is quite similar to --- yet slightly less symmetric than --- the corresponding distribution in the inset of Fig.~2c in the manuscript, while the value of the objective function is very similar to that of $\beta\=4$ ($\eta_{_{3.5}}\!\simeq\!0.36$, see caption, compared to $\eta_{_{4}}\!\simeq\!0.43$, see caption of Fig.~2 in the manuscript). Consequently, we conclude that $\beta\=3.5$ is roughly similarly consistent with the experimental data for glassy glycerol compared to $\beta\=4$.

In the bottom row of Fig.~\ref{fig:figS2}, we present the corresponding analysis for $\beta\=4.5$. It is observed in the inset of Fig.~\ref{fig:figS2}c that the distribution of $\Delta_i$ around its mean is broader than the corresponding distribution in the inset of Fig.~2c in the manuscript, resulting in a larger value of the minimal (optimal) objective function compared to the corrsponding value for $\beta\=4$ ($\eta_{_{4.5}}\!\simeq\!0.61$, see caption, compared to $\eta_{_{4}}\!\simeq\!0.43$, see caption of Fig.~2 in the manuscript). Consequently, we conclude that $\beta\=4.5$ is somewhat less consistent with the experimental data for glassy glycerol compared to $\beta\=4$. Taken together, it appears as if the experimental data in this case is consistent with $\beta\=4$ within 10-15\%. We attribute the larger uncertainty in $\beta$ in the glassy glycerol dataset compared to the glassy dibutyl phthalate dataset to the emerging values of the optimal $\eta_{_{4}}$, $\eta_{_{4}}\!\simeq\!0.43$ and $\eta_{_{4}}\!\simeq\!0.05$, respectively. We note that $\eta_{_{4}}\!\simeq\!0.43$ is an outlier in our analysis (see manuscript) and that in general --- as stressed --- the value of the minimal/optimal $\eta_{_{4}}$ does not exclusively determine the consistency with $\beta\=4$, but rather the three assessment criteria together.\\

\section{T\lowercase{he dimensionless frequency ratio} $A_{\rm g}^{-1/5}\!/\omega_{_{\rm D}}$ \lowercase{in computer glasses}}
\label{sec:frequency_ratio}

In the manuscript, we reported that $A_{\rm g}^{-1/5}/\omega_{_{\rm D}}$ in computer glasses spans the range $0.22-0.73$. Here, we specify the literature sources of this data compilation.

We first consider a polydisperse glass-forming model~\cite{LB_swap_prx} that can be very deeply
supercooled (possibly deeper than laboratory glasses) using the swap-Monte-Carlo method, studied in~\cite{phonon_width_2}. In particular, the results presented in Fig.~1c of~\cite{phonon_width_2} correspond to $A_{\rm g}^{-1/5}/\omega_{_{\rm D}}\!\simeq\!0.22$, obtained for glassy states quenched from a `parent temperature' of $T_{\rm p}\!=\!1.3$ (see~\cite{phonon_width_2} for details). The results presented in Fig.~6a of the same work~\cite{phonon_width_2}, corresponding to significantly deeper supercooling characterized by $T_{\rm p}\!=\!0.32$, gave rise to $A_{\rm g}^{-1/5}/\omega_{_{\rm D}}\!\simeq\!0.73$. Finally, we consider the so-called `sticky spheres' glass-forming model of~\cite{sticky_spheres_part2}. Specifically, the results presented in Fig.~7a of~\cite{sticky_spheres_part2}, corresponding to $r_{\rm c}\!=\!1.2$ and $T_{\rm p}\!=\!0.77$ (see details therein), gave rise to $A_{\rm g}^{-1/5}/\omega_{_{\rm D}}\!\simeq\!0.36$.


\begin{thebibliography}{45}%
\makeatletter
\providecommand \@ifxundefined [1]{%
 \@ifx{#1\undefined}
}%
\providecommand \@ifnum [1]{%
 \ifnum #1\expandafter \@firstoftwo
 \else \expandafter \@secondoftwo
 \fi
}%
\providecommand \@ifx [1]{%
 \ifx #1\expandafter \@firstoftwo
 \else \expandafter \@secondoftwo
 \fi
}%
\providecommand \natexlab [1]{#1}%
\providecommand \enquote  [1]{``#1''}%
\providecommand \bibnamefont  [1]{#1}%
\providecommand \bibfnamefont [1]{#1}%
\providecommand \citenamefont [1]{#1}%
\providecommand \href@noop [0]{\@secondoftwo}%
\providecommand \href [0]{\begingroup \@sanitize@url \@href}%
\providecommand \@href[1]{\@@startlink{#1}\@@href}%
\providecommand \@@href[1]{\endgroup#1\@@endlink}%
\providecommand \@sanitize@url [0]{\catcode `\\12\catcode `\$12\catcode
  `\&12\catcode `\#12\catcode `\^12\catcode `\_12\catcode `\%12\relax}%
\providecommand \@@startlink[1]{}%
\providecommand \@@endlink[0]{}%
\providecommand \url  [0]{\begingroup\@sanitize@url \@url }%
\providecommand \@url [1]{\endgroup\@href {#1}{\urlprefix }}%
\providecommand \urlprefix  [0]{URL }%
\providecommand \Eprint [0]{\href }%
\providecommand \doibase [0]{https://doi.org/}%
\providecommand \selectlanguage [0]{\@gobble}%
\providecommand \bibinfo  [0]{\@secondoftwo}%
\providecommand \bibfield  [0]{\@secondoftwo}%
\providecommand \translation [1]{[#1]}%
\providecommand \BibitemOpen [0]{}%
\providecommand \bibitemStop [0]{}%
\providecommand \bibitemNoStop [0]{.\EOS\space}%
\providecommand \EOS [0]{\spacefactor3000\relax}%
\providecommand \BibitemShut  [1]{\csname bibitem#1\endcsname}%
\let\auto@bib@innerbib\@empty
\bibitem [{\citenamefont {Phillips}(1972)}]{phillips1972tunneling}%
  \BibitemOpen
  \bibfield  {author} {\bibinfo {author} {\bibfnamefont {W.}~\bibnamefont
  {Phillips}},\ }\bibfield  {title} {\bibinfo {title} {Tunneling states in
  amorphous solids},\ }\href {https://doi.org/10.1007/BF00660072} {\bibfield
  {journal} {\bibinfo  {journal} {J. Low Temp. Phys.}\ }\textbf {\bibinfo
  {volume} {7}},\ \bibinfo {pages} {351} (\bibinfo {year} {1972})}\BibitemShut
  {NoStop}%
\bibitem [{\citenamefont {Anderson}\ \emph {et~al.}(1972)\citenamefont
  {Anderson}, \citenamefont {Halperin},\ and\ \citenamefont
  {Varma}}]{anderson1972anomalous}%
  \BibitemOpen
  \bibfield  {author} {\bibinfo {author} {\bibfnamefont {P.~W.}\ \bibnamefont
  {Anderson}}, \bibinfo {author} {\bibfnamefont {B.~I.}\ \bibnamefont
  {Halperin}},\ and\ \bibinfo {author} {\bibfnamefont {C.~M.}\ \bibnamefont
  {Varma}},\ }\bibfield  {title} {\bibinfo {title} {Anomalous low-temperature
  thermal properties of glasses and spin glasses},\ }\href
  {https://doi.org/10.1080/14786437208229210} {\bibfield  {journal} {\bibinfo
  {journal} {Philos. Mag.}\ }\textbf {\bibinfo {volume} {25}},\ \bibinfo
  {pages} {1} (\bibinfo {year} {1972})}\BibitemShut {NoStop}%
\bibitem [{\citenamefont {Buchenau}\ \emph {et~al.}(1991)\citenamefont
  {Buchenau}, \citenamefont {Galperin}, \citenamefont {Gurevich},\ and\
  \citenamefont {Schober}}]{soft_potential_model_1991}%
  \BibitemOpen
  \bibfield  {author} {\bibinfo {author} {\bibfnamefont {U.}~\bibnamefont
  {Buchenau}}, \bibinfo {author} {\bibfnamefont {Y.~M.}\ \bibnamefont
  {Galperin}}, \bibinfo {author} {\bibfnamefont {V.~L.}\ \bibnamefont
  {Gurevich}},\ and\ \bibinfo {author} {\bibfnamefont {H.~R.}\ \bibnamefont
  {Schober}},\ }\bibfield  {title} {\bibinfo {title} {Anharmonic potentials and
  vibrational localization in glasses},\ }\href
  {https://doi.org/10.1103/PhysRevB.43.5039} {\bibfield  {journal} {\bibinfo
  {journal} {Phys. Rev. B}\ }\textbf {\bibinfo {volume} {43}},\ \bibinfo
  {pages} {5039} (\bibinfo {year} {1991})}\BibitemShut {NoStop}%
\bibitem [{\citenamefont {Pohl}\ \emph {et~al.}(2002)\citenamefont {Pohl},
  \citenamefont {Liu},\ and\ \citenamefont {Thompson}}]{pohl_review}%
  \BibitemOpen
  \bibfield  {author} {\bibinfo {author} {\bibfnamefont {R.~O.}\ \bibnamefont
  {Pohl}}, \bibinfo {author} {\bibfnamefont {X.}~\bibnamefont {Liu}},\ and\
  \bibinfo {author} {\bibfnamefont {E.}~\bibnamefont {Thompson}},\ }\bibfield
  {title} {\bibinfo {title} {Low-temperature thermal conductivity and acoustic
  attenuation in amorphous solids},\ }\href
  {https://doi.org/10.1103/RevModPhys.74.991} {\bibfield  {journal} {\bibinfo
  {journal} {Rev. Mod. Phys.}\ }\textbf {\bibinfo {volume} {74}},\ \bibinfo
  {pages} {991} (\bibinfo {year} {2002})}\BibitemShut {NoStop}%
\bibitem [{\citenamefont {Ramos}(2022)}]{ramos2022book}%
  \BibitemOpen
  \bibfield  {author} {\bibinfo {author} {\bibfnamefont {M.~A.}\ \bibnamefont
  {Ramos}},\ }\href {https://doi.org/10.1142/q0371} {\emph {\bibinfo {title}
  {Low-Temperature Thermal and Vibrational Properties of Disordered Solids}}}\
  (\bibinfo  {publisher} {World Scientific},\ \bibinfo {year}
  {2022})\BibitemShut {NoStop}%
\bibitem [{\citenamefont {Chaikin}\ and\ \citenamefont
  {Lubensky}(1995)}]{chaikin_lubensky_book}%
  \BibitemOpen
  \bibfield  {author} {\bibinfo {author} {\bibfnamefont {P.}~\bibnamefont
  {Chaikin}}\ and\ \bibinfo {author} {\bibfnamefont {T.}~\bibnamefont
  {Lubensky}},\ }\href@noop {} {\emph {\bibinfo {title} {Principles of
  Condensed Matter Physics}}}\ (\bibinfo  {publisher} {Cambridge University
  Press},\ \bibinfo {year} {1995})\BibitemShut {NoStop}%
\bibitem [{\citenamefont {Il'in}\ \emph {et~al.}(1987)\citenamefont {Il'in},
  \citenamefont {Karpov},\ and\ \citenamefont
  {Parshin}}]{soft_potential_model_1987}%
  \BibitemOpen
  \bibfield  {author} {\bibinfo {author} {\bibfnamefont {M.}~\bibnamefont
  {Il'in}}, \bibinfo {author} {\bibfnamefont {V.}~\bibnamefont {Karpov}},\ and\
  \bibinfo {author} {\bibfnamefont {D.}~\bibnamefont {Parshin}},\ }\bibfield
  {title} {\bibinfo {title} {Parameters of soft atomic potentials in glasses},\
  }\href {http://jetp.ras.ru/cgi-bin/dn/e_065_01_0165} {\bibfield  {journal}
  {\bibinfo  {journal} {Zh. Eksp. Teor. Fiz.}\ }\textbf {\bibinfo {volume}
  {92}},\ \bibinfo {pages} {291} (\bibinfo {year} {1987})}\BibitemShut
  {NoStop}%
\bibitem [{\citenamefont {Gurevich}\ \emph {et~al.}(2003)\citenamefont
  {Gurevich}, \citenamefont {Parshin},\ and\ \citenamefont
  {Schober}}]{Gurevich2003}%
  \BibitemOpen
  \bibfield  {author} {\bibinfo {author} {\bibfnamefont {V.~L.}\ \bibnamefont
  {Gurevich}}, \bibinfo {author} {\bibfnamefont {D.~A.}\ \bibnamefont
  {Parshin}},\ and\ \bibinfo {author} {\bibfnamefont {H.~R.}\ \bibnamefont
  {Schober}},\ }\bibfield  {title} {\bibinfo {title} {Anharmonicity,
  vibrational instability, and the boson peak in glasses},\ }\href
  {https://doi.org/10.1103/PhysRevB.67.094203} {\bibfield  {journal} {\bibinfo
  {journal} {Phys. Rev. B}\ }\textbf {\bibinfo {volume} {67}},\ \bibinfo
  {pages} {094203} (\bibinfo {year} {2003})}\BibitemShut {NoStop}%
\bibitem [{\citenamefont {Schirmacher}\ \emph {et~al.}(2007)\citenamefont
  {Schirmacher}, \citenamefont {Ruocco},\ and\ \citenamefont
  {Scopigno}}]{Schirmacher_prl_2007}%
  \BibitemOpen
  \bibfield  {author} {\bibinfo {author} {\bibfnamefont {W.}~\bibnamefont
  {Schirmacher}}, \bibinfo {author} {\bibfnamefont {G.}~\bibnamefont
  {Ruocco}},\ and\ \bibinfo {author} {\bibfnamefont {T.}~\bibnamefont
  {Scopigno}},\ }\bibfield  {title} {\bibinfo {title} {Acoustic attenuation in
  glasses and its relation with the boson peak},\ }\href
  {https://doi.org/10.1103/PhysRevLett.98.025501} {\bibfield  {journal}
  {\bibinfo  {journal} {Phys. Rev. Lett.}\ }\textbf {\bibinfo {volume} {98}},\
  \bibinfo {pages} {025501} (\bibinfo {year} {2007})}\BibitemShut {NoStop}%
\bibitem [{\citenamefont {Shintani}\ and\ \citenamefont
  {Tanaka}(2008)}]{tanaka_boson_peak_2008}%
  \BibitemOpen
  \bibfield  {author} {\bibinfo {author} {\bibfnamefont {H.}~\bibnamefont
  {Shintani}}\ and\ \bibinfo {author} {\bibfnamefont {H.}~\bibnamefont
  {Tanaka}},\ }\bibfield  {title} {\bibinfo {title} {Universal link between the
  boson peak and transverse phonons in glass},\ }\href
  {https://doi.org/10.1038/nmat2293} {\bibfield  {journal} {\bibinfo  {journal}
  {Nat. Mater.}\ }\textbf {\bibinfo {volume} {7}},\ \bibinfo {pages} {870}
  (\bibinfo {year} {2008})}\BibitemShut {NoStop}%
\bibitem [{\citenamefont {DeGiuli}\ \emph {et~al.}(2014)\citenamefont
  {DeGiuli}, \citenamefont {Laversanne-Finot}, \citenamefont {During},
  \citenamefont {Lerner},\ and\ \citenamefont {Wyart}}]{eric_boson_peak_emt}%
  \BibitemOpen
  \bibfield  {author} {\bibinfo {author} {\bibfnamefont {E.}~\bibnamefont
  {DeGiuli}}, \bibinfo {author} {\bibfnamefont {A.}~\bibnamefont
  {Laversanne-Finot}}, \bibinfo {author} {\bibfnamefont {G.}~\bibnamefont
  {During}}, \bibinfo {author} {\bibfnamefont {E.}~\bibnamefont {Lerner}},\
  and\ \bibinfo {author} {\bibfnamefont {M.}~\bibnamefont {Wyart}},\ }\bibfield
   {title} {\bibinfo {title} {Effects of coordination and pressure on sound
  attenuation{,} boson peak and elasticity in amorphous solids},\ }\href
  {https://doi.org/10.1039/C4SM00561A} {\bibfield  {journal} {\bibinfo
  {journal} {Soft Matter}\ }\textbf {\bibinfo {volume} {10}},\ \bibinfo {pages}
  {5628} (\bibinfo {year} {2014})}\BibitemShut {NoStop}%
\bibitem [{\citenamefont {Lerner}\ and\ \citenamefont
  {Bouchbinder}(2021)}]{JCP_Perspective}%
  \BibitemOpen
  \bibfield  {author} {\bibinfo {author} {\bibfnamefont {E.}~\bibnamefont
  {Lerner}}\ and\ \bibinfo {author} {\bibfnamefont {E.}~\bibnamefont
  {Bouchbinder}},\ }\bibfield  {title} {\bibinfo {title} {Low-energy
  quasilocalized excitations in structural glasses},\ }\href
  {https://doi.org/10.1063/5.0069477} {\bibfield  {journal} {\bibinfo
  {journal} {J. Chem. Phys.}\ }\textbf {\bibinfo {volume} {155}},\ \bibinfo
  {pages} {200901} (\bibinfo {year} {2021})}\BibitemShut {NoStop}%
\bibitem [{\citenamefont {Lerner}\ \emph {et~al.}(2016)\citenamefont {Lerner},
  \citenamefont {D\"uring},\ and\ \citenamefont
  {Bouchbinder}}]{modes_prl_2016}%
  \BibitemOpen
  \bibfield  {author} {\bibinfo {author} {\bibfnamefont {E.}~\bibnamefont
  {Lerner}}, \bibinfo {author} {\bibfnamefont {G.}~\bibnamefont {D\"uring}},\
  and\ \bibinfo {author} {\bibfnamefont {E.}~\bibnamefont {Bouchbinder}},\
  }\bibfield  {title} {\bibinfo {title} {Statistics and properties of
  low-frequency vibrational modes in structural glasses},\ }\href
  {https://doi.org/10.1103/PhysRevLett.117.035501} {\bibfield  {journal}
  {\bibinfo  {journal} {Phys. Rev. Lett.}\ }\textbf {\bibinfo {volume} {117}},\
  \bibinfo {pages} {035501} (\bibinfo {year} {2016})}\BibitemShut {NoStop}%
\bibitem [{\citenamefont {Mizuno}\ \emph {et~al.}(2017)\citenamefont {Mizuno},
  \citenamefont {Shiba},\ and\ \citenamefont {Ikeda}}]{ikeda_pnas}%
  \BibitemOpen
  \bibfield  {author} {\bibinfo {author} {\bibfnamefont {H.}~\bibnamefont
  {Mizuno}}, \bibinfo {author} {\bibfnamefont {H.}~\bibnamefont {Shiba}},\ and\
  \bibinfo {author} {\bibfnamefont {A.}~\bibnamefont {Ikeda}},\ }\bibfield
  {title} {\bibinfo {title} {Continuum limit of the vibrational properties of
  amorphous solids},\ }\href {https://doi.org/10.1073/pnas.1709015114}
  {\bibfield  {journal} {\bibinfo  {journal} {Proc. Natl. Acad. Sci. U.S.A.}\
  }\textbf {\bibinfo {volume} {114}},\ \bibinfo {pages} {E9767} (\bibinfo
  {year} {2017})}\BibitemShut {NoStop}%
\bibitem [{\citenamefont {Kapteijns}\ \emph {et~al.}(2018)\citenamefont
  {Kapteijns}, \citenamefont {Bouchbinder},\ and\ \citenamefont
  {Lerner}}]{modes_prl_2018}%
  \BibitemOpen
  \bibfield  {author} {\bibinfo {author} {\bibfnamefont {G.}~\bibnamefont
  {Kapteijns}}, \bibinfo {author} {\bibfnamefont {E.}~\bibnamefont
  {Bouchbinder}},\ and\ \bibinfo {author} {\bibfnamefont {E.}~\bibnamefont
  {Lerner}},\ }\bibfield  {title} {\bibinfo {title} {Universal nonphononic
  density of states in \uppercase{2D, 3D}, and \uppercase{4D} glasses},\ }\href
  {https://doi.org/10.1103/PhysRevLett.121.055501} {\bibfield  {journal}
  {\bibinfo  {journal} {Phys. Rev. Lett.}\ }\textbf {\bibinfo {volume} {121}},\
  \bibinfo {pages} {055501} (\bibinfo {year} {2018})}\BibitemShut {NoStop}%
\bibitem [{\citenamefont {Wang}\ \emph {et~al.}(2019)\citenamefont {Wang},
  \citenamefont {Ninarello}, \citenamefont {Guan}, \citenamefont {Berthier},
  \citenamefont {Szamel},\ and\ \citenamefont {Flenner}}]{LB_modes_2019}%
  \BibitemOpen
  \bibfield  {author} {\bibinfo {author} {\bibfnamefont {L.}~\bibnamefont
  {Wang}}, \bibinfo {author} {\bibfnamefont {A.}~\bibnamefont {Ninarello}},
  \bibinfo {author} {\bibfnamefont {P.}~\bibnamefont {Guan}}, \bibinfo {author}
  {\bibfnamefont {L.}~\bibnamefont {Berthier}}, \bibinfo {author}
  {\bibfnamefont {G.}~\bibnamefont {Szamel}},\ and\ \bibinfo {author}
  {\bibfnamefont {E.}~\bibnamefont {Flenner}},\ }\bibfield  {title} {\bibinfo
  {title} {Low-frequency vibrational modes of stable glasses},\ }\href
  {https://doi.org/10.1038/s41467-018-07978-1} {\bibfield  {journal} {\bibinfo
  {journal} {Nat. Commun.}\ }\textbf {\bibinfo {volume} {10}},\ \bibinfo
  {pages} {26} (\bibinfo {year} {2019})}\BibitemShut {NoStop}%
\bibitem [{\citenamefont {Rainone}\ \emph {et~al.}(2020)\citenamefont
  {Rainone}, \citenamefont {Bouchbinder},\ and\ \citenamefont
  {Lerner}}]{pinching_pnas}%
  \BibitemOpen
  \bibfield  {author} {\bibinfo {author} {\bibfnamefont {C.}~\bibnamefont
  {Rainone}}, \bibinfo {author} {\bibfnamefont {E.}~\bibnamefont
  {Bouchbinder}},\ and\ \bibinfo {author} {\bibfnamefont {E.}~\bibnamefont
  {Lerner}},\ }\bibfield  {title} {\bibinfo {title} {Pinching a glass reveals
  key properties of its soft spots},\ }\href
  {https://doi.org/10.1073/pnas.1919958117} {\bibfield  {journal} {\bibinfo
  {journal} {Proc. Natl. Acad. Sci. U.S.A.}\ }\textbf {\bibinfo {volume}
  {117}},\ \bibinfo {pages} {5228} (\bibinfo {year} {2020})}\BibitemShut
  {NoStop}%
\bibitem [{\citenamefont {Richard}\ \emph {et~al.}(2020)\citenamefont
  {Richard}, \citenamefont {Gonz\'alez-L\'opez}, \citenamefont {Kapteijns},
  \citenamefont {Pater}, \citenamefont {Vaknin}, \citenamefont {Bouchbinder},\
  and\ \citenamefont {Lerner}}]{modes_prl_2020}%
  \BibitemOpen
  \bibfield  {author} {\bibinfo {author} {\bibfnamefont {D.}~\bibnamefont
  {Richard}}, \bibinfo {author} {\bibfnamefont {K.}~\bibnamefont
  {Gonz\'alez-L\'opez}}, \bibinfo {author} {\bibfnamefont {G.}~\bibnamefont
  {Kapteijns}}, \bibinfo {author} {\bibfnamefont {R.}~\bibnamefont {Pater}},
  \bibinfo {author} {\bibfnamefont {T.}~\bibnamefont {Vaknin}}, \bibinfo
  {author} {\bibfnamefont {E.}~\bibnamefont {Bouchbinder}},\ and\ \bibinfo
  {author} {\bibfnamefont {E.}~\bibnamefont {Lerner}},\ }\bibfield  {title}
  {\bibinfo {title} {Universality of the nonphononic vibrational spectrum
  across different classes of computer glasses},\ }\href
  {https://doi.org/10.1103/PhysRevLett.125.085502} {\bibfield  {journal}
  {\bibinfo  {journal} {Phys. Rev. Lett.}\ }\textbf {\bibinfo {volume} {125}},\
  \bibinfo {pages} {085502} (\bibinfo {year} {2020})}\BibitemShut {NoStop}%
\bibitem [{\citenamefont {Bonfanti}\ \emph {et~al.}(2020)\citenamefont
  {Bonfanti}, \citenamefont {Guerra}, \citenamefont {Mondal}, \citenamefont
  {Procaccia},\ and\ \citenamefont {Zapperi}}]{universal_VDoS_ip}%
  \BibitemOpen
  \bibfield  {author} {\bibinfo {author} {\bibfnamefont {S.}~\bibnamefont
  {Bonfanti}}, \bibinfo {author} {\bibfnamefont {R.}~\bibnamefont {Guerra}},
  \bibinfo {author} {\bibfnamefont {C.}~\bibnamefont {Mondal}}, \bibinfo
  {author} {\bibfnamefont {I.}~\bibnamefont {Procaccia}},\ and\ \bibinfo
  {author} {\bibfnamefont {S.}~\bibnamefont {Zapperi}},\ }\bibfield  {title}
  {\bibinfo {title} {Universal low-frequency vibrational modes in silica
  glasses},\ }\href {https://doi.org/10.1103/PhysRevLett.125.085501} {\bibfield
   {journal} {\bibinfo  {journal} {Phys. Rev. Lett.}\ }\textbf {\bibinfo
  {volume} {125}},\ \bibinfo {pages} {085501} (\bibinfo {year}
  {2020})}\BibitemShut {NoStop}%
\bibitem [{\citenamefont {Lerner}\ and\ \citenamefont
  {Bouchbinder}(2022)}]{2d_spectra_jcp_2022}%
  \BibitemOpen
  \bibfield  {author} {\bibinfo {author} {\bibfnamefont {E.}~\bibnamefont
  {Lerner}}\ and\ \bibinfo {author} {\bibfnamefont {E.}~\bibnamefont
  {Bouchbinder}},\ }\bibfield  {title} {\bibinfo {title} {{Nonphononic spectrum
  of two-dimensional structural glasses}},\ }\href
  {https://doi.org/10.1063/5.0120115} {\bibfield  {journal} {\bibinfo
  {journal} {J. Chem. Phys.}\ }\textbf {\bibinfo {volume} {157}},\ \bibinfo
  {pages} {166101} (\bibinfo {year} {2022})}\BibitemShut {NoStop}%
\bibitem [{\citenamefont {Shiraishi}\ \emph {et~al.}(2023)\citenamefont
  {Shiraishi}, \citenamefont {Mizuno},\ and\ \citenamefont
  {Ikeda}}]{atsushi_2d_pinning_jcp_2023}%
  \BibitemOpen
  \bibfield  {author} {\bibinfo {author} {\bibfnamefont {K.}~\bibnamefont
  {Shiraishi}}, \bibinfo {author} {\bibfnamefont {H.}~\bibnamefont {Mizuno}},\
  and\ \bibinfo {author} {\bibfnamefont {A.}~\bibnamefont {Ikeda}},\ }\bibfield
   {title} {\bibinfo {title} {{Non-phononic density of states of
  two-dimensional glasses revealed by random pinning}},\ }\href
  {https://doi.org/10.1063/5.0142648} {\bibfield  {journal} {\bibinfo
  {journal} {J. Chem. Phys.}\ }\textbf {\bibinfo {volume} {158}},\ \bibinfo
  {pages} {174502} (\bibinfo {year} {2023})}\BibitemShut {NoStop}%
\bibitem [{\citenamefont {Bouchbinder}\ and\ \citenamefont
  {Lerner}(2018)}]{phonon_widths}%
  \BibitemOpen
  \bibfield  {author} {\bibinfo {author} {\bibfnamefont {E.}~\bibnamefont
  {Bouchbinder}}\ and\ \bibinfo {author} {\bibfnamefont {E.}~\bibnamefont
  {Lerner}},\ }\bibfield  {title} {\bibinfo {title} {Universal disorder-induced
  broadening of phonon bands: from disordered lattices to glasses},\ }\href
  {http://stacks.iop.org/1367-2630/20/i=7/a=073022} {\bibfield  {journal}
  {\bibinfo  {journal} {New J. Phys.}\ }\textbf {\bibinfo {volume} {20}},\
  \bibinfo {pages} {073022} (\bibinfo {year} {2018})}\BibitemShut {NoStop}%
\bibitem [{\citenamefont {Lerner}\ and\ \citenamefont
  {Bouchbinder}(2023)}]{boson_peak_2d_jcp_2023}%
  \BibitemOpen
  \bibfield  {author} {\bibinfo {author} {\bibfnamefont {E.}~\bibnamefont
  {Lerner}}\ and\ \bibinfo {author} {\bibfnamefont {E.}~\bibnamefont
  {Bouchbinder}},\ }\bibfield  {title} {\bibinfo {title} {{Boson-peak
  vibrational modes in glasses feature hybridized phononic and quasilocalized
  excitations}},\ }\href {https://doi.org/10.1063/5.0147889} {\bibfield
  {journal} {\bibinfo  {journal} {J. Chem. Phys.}\ }\textbf {\bibinfo {volume}
  {158}},\ \bibinfo {pages} {194503} (\bibinfo {year} {2023})}\BibitemShut
  {NoStop}%
\bibitem [{\citenamefont {Gartner}\ and\ \citenamefont
  {Lerner}(2016)}]{SciPost2016}%
  \BibitemOpen
  \bibfield  {author} {\bibinfo {author} {\bibfnamefont {L.}~\bibnamefont
  {Gartner}}\ and\ \bibinfo {author} {\bibfnamefont {E.}~\bibnamefont
  {Lerner}},\ }\bibfield  {title} {\bibinfo {title} {{Nonlinear modes
  disentangle glassy and Goldstone modes in structural glasses}},\ }\href
  {https://doi.org/10.21468/SciPostPhys.1.2.016} {\bibfield  {journal}
  {\bibinfo  {journal} {SciPost Phys.}\ }\textbf {\bibinfo {volume} {1}},\
  \bibinfo {pages} {016} (\bibinfo {year} {2016})}\BibitemShut {NoStop}%
\bibitem [{\citenamefont {Kapteijns}\ \emph {et~al.}(2020)\citenamefont
  {Kapteijns}, \citenamefont {Richard},\ and\ \citenamefont
  {Lerner}}]{episode_1_2020}%
  \BibitemOpen
  \bibfield  {author} {\bibinfo {author} {\bibfnamefont {G.}~\bibnamefont
  {Kapteijns}}, \bibinfo {author} {\bibfnamefont {D.}~\bibnamefont {Richard}},\
  and\ \bibinfo {author} {\bibfnamefont {E.}~\bibnamefont {Lerner}},\
  }\bibfield  {title} {\bibinfo {title} {Nonlinear quasilocalized excitations
  in glasses: True representatives of soft spots},\ }\href
  {https://doi.org/10.1103/PhysRevE.101.032130} {\bibfield  {journal} {\bibinfo
   {journal} {Phys. Rev. E}\ }\textbf {\bibinfo {volume} {101}},\ \bibinfo
  {pages} {032130} (\bibinfo {year} {2020})}\BibitemShut {NoStop}%
\bibitem [{\citenamefont {Richard}\ \emph {et~al.}(2021)\citenamefont
  {Richard}, \citenamefont {Kapteijns}, \citenamefont {Giannini}, \citenamefont
  {Manning},\ and\ \citenamefont {Lerner}}]{pseudo_harmonic_prl}%
  \BibitemOpen
  \bibfield  {author} {\bibinfo {author} {\bibfnamefont {D.}~\bibnamefont
  {Richard}}, \bibinfo {author} {\bibfnamefont {G.}~\bibnamefont {Kapteijns}},
  \bibinfo {author} {\bibfnamefont {J.~A.}\ \bibnamefont {Giannini}}, \bibinfo
  {author} {\bibfnamefont {M.~L.}\ \bibnamefont {Manning}},\ and\ \bibinfo
  {author} {\bibfnamefont {E.}~\bibnamefont {Lerner}},\ }\bibfield  {title}
  {\bibinfo {title} {Simple and broadly applicable definition of shear
  transformation zones},\ }\href
  {https://doi.org/10.1103/PhysRevLett.126.015501} {\bibfield  {journal}
  {\bibinfo  {journal} {Phys. Rev. Lett.}\ }\textbf {\bibinfo {volume} {126}},\
  \bibinfo {pages} {015501} (\bibinfo {year} {2021})}\BibitemShut {NoStop}%
\bibitem [{\citenamefont {Richard}\ \emph {et~al.}(2023)\citenamefont
  {Richard}, \citenamefont {Kapteijns},\ and\ \citenamefont
  {Lerner}}]{david_detecting_qles_pre_2023}%
  \BibitemOpen
  \bibfield  {author} {\bibinfo {author} {\bibfnamefont {D.}~\bibnamefont
  {Richard}}, \bibinfo {author} {\bibfnamefont {G.}~\bibnamefont {Kapteijns}},\
  and\ \bibinfo {author} {\bibfnamefont {E.}~\bibnamefont {Lerner}},\
  }\bibfield  {title} {\bibinfo {title} {Detecting low-energy quasilocalized
  excitations in computer glasses},\ }\href
  {https://doi.org/10.1103/PhysRevE.108.044124} {\bibfield  {journal} {\bibinfo
   {journal} {Phys. Rev. E}\ }\textbf {\bibinfo {volume} {108}},\ \bibinfo
  {pages} {044124} (\bibinfo {year} {2023})}\BibitemShut {NoStop}%
\bibitem [{\citenamefont {Ramos}(2004)}]{ramos_2004}%
  \BibitemOpen
  \bibfield  {author} {\bibinfo {author} {\bibfnamefont {M.~A.}\ \bibnamefont
  {Ramos}},\ }\bibfield  {title} {\bibinfo {title} {Are the calorimetric and
  elastic debye temperatures of glasses really different?},\ }\href
  {https://doi.org/10.1080/14786430310001644053} {\bibfield  {journal}
  {\bibinfo  {journal} {Philos. Mag.}\ }\textbf {\bibinfo {volume} {84}},\
  \bibinfo {pages} {1313} (\bibinfo {year} {2004})}\BibitemShut {NoStop}%
\bibitem [{\citenamefont {Lerner}\ \emph {et~al.}(2024)\citenamefont {Lerner},
  \citenamefont {Moriel},\ and\ \citenamefont
  {Bouchbinder}}]{additive_structure_2024_JCP}%
  \BibitemOpen
  \bibfield  {author} {\bibinfo {author} {\bibfnamefont {E.}~\bibnamefont
  {Lerner}}, \bibinfo {author} {\bibfnamefont {A.}~\bibnamefont {Moriel}},\
  and\ \bibinfo {author} {\bibfnamefont {E.}~\bibnamefont {Bouchbinder}},\
  }\bibfield  {title} {\bibinfo {title} {Enumerating low-frequency nonphononic
  vibrations in computer glasses},\ }\href {https://doi.org/10.1063/5.0216351}
  {\bibfield  {journal} {\bibinfo  {journal} {J. Chem. Phys}\
  } \textbf {\bibinfo
  {volume} {161}},\ \bibinfo {pages} {014504} (\bibinfo {year}
  {2024})}\BibitemShut {NoStop}%
\bibitem [{\citenamefont {Yannopoulos}\ \emph {et~al.}(2006)\citenamefont
  {Yannopoulos}, \citenamefont {Andrikopoulos},\ and\ \citenamefont
  {Ruocco}}]{YANNOPOULOS20064541}%
  \BibitemOpen
  \bibfield  {author} {\bibinfo {author} {\bibfnamefont {S.}~\bibnamefont
  {Yannopoulos}}, \bibinfo {author} {\bibfnamefont {K.}~\bibnamefont
  {Andrikopoulos}},\ and\ \bibinfo {author} {\bibfnamefont {G.}~\bibnamefont
  {Ruocco}},\ }\bibfield  {title} {\bibinfo {title} {On the analysis of the
  vibrational boson peak and low-energy excitations in glasses},\ }\href
  {https://doi.org/https://doi.org/10.1016/j.jnoncrysol.2006.02.164} {\bibfield
   {journal} {\bibinfo  {journal} {J. Non-Cryst. Solids}\ }\textbf {\bibinfo
  {volume} {352}},\ \bibinfo {pages} {4541} (\bibinfo {year}
  {2006})}\BibitemShut {NoStop}%
\bibitem [{\citenamefont {Kalampounias}\ \emph {et~al.}(2006)\citenamefont
  {Kalampounias}, \citenamefont {Yannopoulos},\ and\ \citenamefont
  {Papatheodorou}}]{KALAMPOUNIAS20064619}%
  \BibitemOpen
  \bibfield  {author} {\bibinfo {author} {\bibfnamefont {A.}~\bibnamefont
  {Kalampounias}}, \bibinfo {author} {\bibfnamefont {S.}~\bibnamefont
  {Yannopoulos}},\ and\ \bibinfo {author} {\bibfnamefont {G.}~\bibnamefont
  {Papatheodorou}},\ }\bibfield  {title} {\bibinfo {title} {A low-frequency
  raman study of glassy, supercooled and molten silica and the preservation of
  the boson peak in the equilibrium liquid state},\ }\href
  {https://doi.org/https://doi.org/10.1016/j.jnoncrysol.2006.02.163} {\bibfield
   {journal} {\bibinfo  {journal} {J. Non-Cryst. Solids}\ }\textbf {\bibinfo
  {volume} {352}},\ \bibinfo {pages} {4619} (\bibinfo {year}
  {2006})}\BibitemShut {NoStop}%
\bibitem [{\citenamefont {Wang}\ \emph {et~al.}(2021)\citenamefont {Wang},
  \citenamefont {Szamel},\ and\ \citenamefont
  {Flenner}}]{grzegorz_2d_modes_prl_2021}%
  \BibitemOpen
  \bibfield  {author} {\bibinfo {author} {\bibfnamefont {L.}~\bibnamefont
  {Wang}}, \bibinfo {author} {\bibfnamefont {G.}~\bibnamefont {Szamel}},\ and\
  \bibinfo {author} {\bibfnamefont {E.}~\bibnamefont {Flenner}},\ }\bibfield
  {title} {\bibinfo {title} {Low-frequency excess vibrational modes in
  two-dimensional glasses},\ }\href
  {https://doi.org/10.1103/PhysRevLett.127.248001} {\bibfield  {journal}
  {\bibinfo  {journal} {Phys. Rev. Lett.}\ }\textbf {\bibinfo {volume} {127}},\
  \bibinfo {pages} {248001} (\bibinfo {year} {2021})}\BibitemShut {NoStop}%
\bibitem [{foo({\natexlab{a}})}]{footnote}%
  \BibitemOpen
  \bibinfo {note} {$A_{\rm D}\!\equiv\!3/\omega_{_{\rm D}}^3$, where Debye's
  frequency is given by $\omega_{_{\rm D}}\!\equiv\!\left(\!
  \frac{18\pi^2(N/V)}{2c_{\rm s}^{-3}+c_\ell^{-3}}\!\right)^{\!1\!/3}$, with
  $N$ denoting the number of particles, $V$ the volume, and $c_{\rm s,\ell}$
  the shear and longitudinal wave-speeds, respectively.}\BibitemShut {Stop}%
\bibitem [{\citenamefont {Moriel}\ \emph {et~al.}(2024)\citenamefont {Moriel},
  \citenamefont {Lerner},\ and\ \citenamefont {Bouchbinder}}]{moriel2024boson}%
  \BibitemOpen
  \bibfield  {author} {\bibinfo {author} {\bibfnamefont {A.}~\bibnamefont
  {Moriel}}, \bibinfo {author} {\bibfnamefont {E.}~\bibnamefont {Lerner}},\
  and\ \bibinfo {author} {\bibfnamefont {E.}~\bibnamefont {Bouchbinder}},\
  }\bibfield  {title} {\bibinfo {title} {Boson peak in the vibrational spectra
  of glasses},\ }\href {https://doi.org/10.1103/PhysRevResearch.6.023053}
  {\bibfield  {journal} {\bibinfo  {journal} {Phys. Rev. Res.}\ }\textbf
  {\bibinfo {volume} {6}},\ \bibinfo {pages} {023053} (\bibinfo {year}
  {2024})}\BibitemShut {NoStop}%
\bibitem [{\citenamefont {Buchenau}(1999)}]{buchenau1999neutron}%
  \BibitemOpen
  \bibfield  {author} {\bibinfo {author} {\bibfnamefont {U.}~\bibnamefont
  {Buchenau}},\ }\bibfield  {title} {\bibinfo {title} {Neutron and x-ray
  scattering from glasses},\ }\href {https://doi.org/10.1063/1.1301447}
  {\bibfield  {journal} {\bibinfo  {journal} {AIP Conf. Proc.}\ }\textbf
  {\bibinfo {volume} {489}},\ \bibinfo {pages} {3–23} (\bibinfo {year}
  {1999})}\BibitemShut {NoStop}%
\bibitem [{\citenamefont {Taraskin}\ and\ \citenamefont
  {Elliott}(1997)}]{taraskin1997connection}%
  \BibitemOpen
  \bibfield  {author} {\bibinfo {author} {\bibfnamefont {S.}~\bibnamefont
  {Taraskin}}\ and\ \bibinfo {author} {\bibfnamefont {S.}~\bibnamefont
  {Elliott}},\ }\bibfield  {title} {\bibinfo {title} {Connection between the
  true vibrational density of states and that derived from inelastic neutron
  scattering},\ }\href {https://doi.org/10.1103/PhysRevB.55.117} {\bibfield
  {journal} {\bibinfo  {journal} {Phys. Rev. B}\ }\textbf {\bibinfo {volume}
  {55}},\ \bibinfo {pages} {117} (\bibinfo {year} {1997})}\BibitemShut
  {NoStop}%
\bibitem [{foo({\natexlab{b}})}]{footnote_Matlab}%
  \BibitemOpen
  \bibinfo {note} {Using Matlab's minimization function
  `fminsearch'~\cite{matlab}.}\BibitemShut {Stop}%
\bibitem [{\citenamefont {Chumakov}\ \emph {et~al.}(2004)\citenamefont
  {Chumakov}, \citenamefont {Sergueev}, \citenamefont {Van~B{\"u}rck},
  \citenamefont {Schirmacher}, \citenamefont {Asthalter}, \citenamefont
  {R{\"u}ffer}, \citenamefont {Leupold},\ and\ \citenamefont
  {Petry}}]{chumakov2004collective}%
  \BibitemOpen
  \bibfield  {author} {\bibinfo {author} {\bibfnamefont {A.}~\bibnamefont
  {Chumakov}}, \bibinfo {author} {\bibfnamefont {I.}~\bibnamefont {Sergueev}},
  \bibinfo {author} {\bibfnamefont {U.}~\bibnamefont {Van~B{\"u}rck}}, \bibinfo
  {author} {\bibfnamefont {W.}~\bibnamefont {Schirmacher}}, \bibinfo {author}
  {\bibfnamefont {T.}~\bibnamefont {Asthalter}}, \bibinfo {author}
  {\bibfnamefont {R.}~\bibnamefont {R{\"u}ffer}}, \bibinfo {author}
  {\bibfnamefont {O.}~\bibnamefont {Leupold}},\ and\ \bibinfo {author}
  {\bibfnamefont {W.}~\bibnamefont {Petry}},\ }\bibfield  {title} {\bibinfo
  {title} {Collective nature of the boson peak and universal transboson
  dynamics of glasses},\ }\href {https://doi.org/10.1103/PhysRevLett.92.245508}
  {\bibfield  {journal} {\bibinfo  {journal} {Phys. Rev. Lett.}\ }\textbf
  {\bibinfo {volume} {92}},\ \bibinfo {pages} {245508} (\bibinfo {year}
  {2004})}\BibitemShut {NoStop}%
\bibitem [{\citenamefont {Zorn}\ \emph {et~al.}(1995)\citenamefont {Zorn},
  \citenamefont {Arbe}, \citenamefont {Colmenero}, \citenamefont {Frick},
  \citenamefont {Richter},\ and\ \citenamefont {Buchenau}}]{zorn1995neutron}%
  \BibitemOpen
  \bibfield  {author} {\bibinfo {author} {\bibfnamefont {R.}~\bibnamefont
  {Zorn}}, \bibinfo {author} {\bibfnamefont {A.}~\bibnamefont {Arbe}}, \bibinfo
  {author} {\bibfnamefont {J.}~\bibnamefont {Colmenero}}, \bibinfo {author}
  {\bibfnamefont {B.}~\bibnamefont {Frick}}, \bibinfo {author} {\bibfnamefont
  {D.}~\bibnamefont {Richter}},\ and\ \bibinfo {author} {\bibfnamefont
  {U.}~\bibnamefont {Buchenau}},\ }\bibfield  {title} {\bibinfo {title}
  {Neutron scattering study of the picosecond dynamics of polybutadiene and
  polyisoprene},\ }\href {https://doi.org/10.1103/PhysRevE.52.781} {\bibfield
  {journal} {\bibinfo  {journal} {Phys. Rev. E}\ }\textbf {\bibinfo {volume}
  {52}},\ \bibinfo {pages} {781} (\bibinfo {year} {1995})}\BibitemShut
  {NoStop}%
\bibitem [{\citenamefont {Wuttke}\ \emph {et~al.}(1995)\citenamefont {Wuttke},
  \citenamefont {Petry}, \citenamefont {Coddens},\ and\ \citenamefont
  {Fujara}}]{wuttke1995fast}%
  \BibitemOpen
  \bibfield  {author} {\bibinfo {author} {\bibfnamefont {J.}~\bibnamefont
  {Wuttke}}, \bibinfo {author} {\bibfnamefont {W.}~\bibnamefont {Petry}},
  \bibinfo {author} {\bibfnamefont {G.}~\bibnamefont {Coddens}},\ and\ \bibinfo
  {author} {\bibfnamefont {F.}~\bibnamefont {Fujara}},\ }\bibfield  {title}
  {\bibinfo {title} {Fast dynamics of glass-forming glycerol},\ }\href
  {https://doi.org/10.1103/PhysRevE.52.4026} {\bibfield  {journal} {\bibinfo
  {journal} {Phys. Rev. E}\ }\textbf {\bibinfo {volume} {52}},\ \bibinfo
  {pages} {4026} (\bibinfo {year} {1995})}\BibitemShut {NoStop}%
\bibitem [{\citenamefont {Chumakov}\ \emph {et~al.}(2014)\citenamefont
  {Chumakov}, \citenamefont {Monaco}, \citenamefont {Fontana}, \citenamefont
  {Bosak}, \citenamefont {Hermann}, \citenamefont {Bessas}, \citenamefont
  {Wehinger}, \citenamefont {Crichton}, \citenamefont {Krisch}, \citenamefont
  {R{\"u}ffer} \emph {et~al.}}]{chumakov2014role}%
  \BibitemOpen
  \bibfield  {author} {\bibinfo {author} {\bibfnamefont {A.~I.}\ \bibnamefont
  {Chumakov}}, \bibinfo {author} {\bibfnamefont {G.}~\bibnamefont {Monaco}},
  \bibinfo {author} {\bibfnamefont {A.}~\bibnamefont {Fontana}}, \bibinfo
  {author} {\bibfnamefont {A.}~\bibnamefont {Bosak}}, \bibinfo {author}
  {\bibfnamefont {R.~P.}\ \bibnamefont {Hermann}}, \bibinfo {author}
  {\bibfnamefont {D.}~\bibnamefont {Bessas}}, \bibinfo {author} {\bibfnamefont
  {B.}~\bibnamefont {Wehinger}}, \bibinfo {author} {\bibfnamefont {W.~A.}\
  \bibnamefont {Crichton}}, \bibinfo {author} {\bibfnamefont {M.}~\bibnamefont
  {Krisch}}, \bibinfo {author} {\bibfnamefont {R.}~\bibnamefont {R{\"u}ffer}},
  \emph {et~al.},\ }\bibfield  {title} {\bibinfo {title} {Role of disorder in
  the thermodynamics and atomic dynamics of glasses},\ }\href
  {https://doi.org/10.1103/PhysRevLett.112.025502} {\bibfield  {journal}
  {\bibinfo  {journal} {Phys. Rev. Lett.}\ }\textbf {\bibinfo {volume} {112}},\
  \bibinfo {pages} {025502} (\bibinfo {year} {2014})}\BibitemShut {NoStop}%
\bibitem [{\citenamefont {Rohatgi}(2024)}]{WebPlotDig}%
  \BibitemOpen
  \bibfield  {author} {\bibinfo {author} {\bibfnamefont {A.}~\bibnamefont
  {Rohatgi}},\ }\href {https://automeris.io/WebPlotDigitizer.html} {\bibinfo
  {title} {Webplotdigitizer}} (\bibinfo {year} {2024})\BibitemShut {NoStop}%
\bibitem [{SM()}]{SM}%
  \BibitemOpen
  \href@noop {} {\bibinfo  {journal} {See Supplemental Materials file, which includes
  Refs.~\cite{LB_swap_prx,phonon_width_2,sticky_spheres_part2}, for additional information}\ }\BibitemShut
  {NoStop}%
\bibitem [{\citenamefont {{The MathWorks Inc.}}(2022)}]{matlab}%
  \BibitemOpen
\bibfield  {journal} {  }\bibfield  {author} {\bibinfo {author} {\bibnamefont
  {{The MathWorks Inc.}}},\ }\href {https://www.mathworks.com} {\bibinfo
  {title} {Matlab version: 9.13.0.2166757 ({R2022b})}} (\bibinfo {year}
  {2022})\BibitemShut {NoStop}%
\bibitem [{\citenamefont {Ninarello}\ \emph {et~al.}(2017)\citenamefont
  {Ninarello}, \citenamefont {Berthier},\ and\ \citenamefont
  {Coslovich}}]{LB_swap_prx}%
  \BibitemOpen
  \bibfield  {author} {\bibinfo {author} {\bibfnamefont {A.}~\bibnamefont
  {Ninarello}}, \bibinfo {author} {\bibfnamefont {L.}~\bibnamefont
  {Berthier}},\ and\ \bibinfo {author} {\bibfnamefont {D.}~\bibnamefont
  {Coslovich}},\ }\bibfield  {title} {\bibinfo {title} {Models and algorithms
  for the next generation of glass transition studies},\ }\href
  {https://doi.org/10.1103/PhysRevX.7.021039} {\bibfield  {journal} {\bibinfo
  {journal} {Phys. Rev. X}\ }\textbf {\bibinfo {volume} {7}},\ \bibinfo {pages}
  {021039} (\bibinfo {year} {2017})}\BibitemShut {NoStop}%
\bibitem [{\citenamefont {Kapteijns}\ \emph {et~al.}(2021)\citenamefont
  {Kapteijns}, \citenamefont {Bouchbinder},\ and\ \citenamefont
  {Lerner}}]{phonon_width_2}%
  \BibitemOpen
  \bibfield  {author} {\bibinfo {author} {\bibfnamefont {G.}~\bibnamefont
  {Kapteijns}}, \bibinfo {author} {\bibfnamefont {E.}~\bibnamefont
  {Bouchbinder}},\ and\ \bibinfo {author} {\bibfnamefont {E.}~\bibnamefont
  {Lerner}},\ }\bibfield  {title} {\bibinfo {title} {Unified quantifier of
  mechanical disorder in solids},\ }\href
  {https://doi.org/10.1103/PhysRevE.104.035001} {\bibfield  {journal} {\bibinfo
   {journal} {Phys. Rev. E}\ }\textbf {\bibinfo {volume} {104}},\ \bibinfo
  {pages} {035001} (\bibinfo {year} {2021})}\BibitemShut {NoStop}%
\bibitem [{\citenamefont {Gonz\'alez-L\'opez}\ \emph
  {et~al.}(2021)\citenamefont {Gonz\'alez-L\'opez}, \citenamefont {Shivam},
  \citenamefont {Zheng}, \citenamefont {Ciamarra},\ and\ \citenamefont
  {Lerner}}]{sticky_spheres_part2}%
  \BibitemOpen
  \bibfield  {author} {\bibinfo {author} {\bibfnamefont {K.}~\bibnamefont
  {Gonz\'alez-L\'opez}}, \bibinfo {author} {\bibfnamefont {M.}~\bibnamefont
  {Shivam}}, \bibinfo {author} {\bibfnamefont {Y.}~\bibnamefont {Zheng}},
  \bibinfo {author} {\bibfnamefont {M.~P.}\ \bibnamefont {Ciamarra}},\ and\
  \bibinfo {author} {\bibfnamefont {E.}~\bibnamefont {Lerner}},\ }\bibfield
  {title} {\bibinfo {title} {Mechanical disorder of sticky-sphere glasses.
  \uppercase{II. T}hermomechanical inannealability},\ }\href
  {https://doi.org/10.1103/PhysRevE.103.022606} {\bibfield  {journal} {\bibinfo
   {journal} {Phys. Rev. E}\ }\textbf {\bibinfo {volume} {103}},\ \bibinfo
  {pages} {022606} (\bibinfo {year} {2021})}\BibitemShut {NoStop}%
\end{thebibliography}

%

\end{document}